\documentclass[a4paper]{article}
\pdfoutput=1
\usepackage{amssymb,amsmath}
\usepackage{amsfonts}
\usepackage{hyperref}
\usepackage{subfigure}
\usepackage[utf8]{inputenc}
\usepackage[T1]{fontenc}
\usepackage{lmodern}
\usepackage{tikz}
\usepackage{units}
\usepackage[numbers,sort&compress]{natbib}

\def\Eq#1{Eq.~(\ref{eq-#1})}
\def\Eqs#1#2{Eqs.~(\ref{eq-#1}-\ref{eq-#2})}
\def\Fig#1{Fig.~\ref{fig-#1}}
\def\Sect#1{Sect.~\ref{sec-#1}}
\def\citep#1{\cite{#1}}

\begin{document}

\title{\bf Driving mechanisms and streamwise homogeneity in molecular dynamics simulations of nanochannel flows}
\author{~\\ Vicente Bitrián$^{1,}$\thanks{vicente.bitrian@upc.edu} { and} Javier Principe$^{1,2,}$\thanks{javier.principe@upc.edu}
~\\~\\
{\small $^1$ Departament de Mecànica de Fluids, Universitat Politècnica de Catalunya,} \\
{\small Eduard Maristany, 10-14, 08019, Barcelona, Spain.}
~\\~\\
{\small $^2$ Centre Internacional de Mètodes Numèrics en Enginyeria, } \\
{\small Parc Mediterrani de la Tecnologia, Esteve Terrades 5, 08860, Castelldefels, Spain.}\\
}
\date{\small \today}

\maketitle

\begin{abstract}
In molecular dynamics simulations, nanochannel flows are usually driven by a constant force, that aims to represent a pressure difference between inlet and outlet, and periodic boundary conditions are applied in the streamwise direction resulting in an homogeneous flow. The homogeneity hypothesis can be eliminated adding reservoirs at the inlet and outlet of the channel which permits to predict streamwise variation of flow properties. It also opens the door to drive the flow by applying pressure gradient instead of a constant force. We analyze the impact of these modeling modifications in the prediction of the flow properties and we show when they make a difference with respect to the standard approach. 
It turns out that both assumptions are irrelevant when low pressure differences are considered, but important differences are observed at high pressure differences. They include the density and velocity variation along the channel (the mass flow rate is constant) but, more importantly, the temperature increase and slip length decrease. Because viscous heating is important at high shear rates, these modeling issues are also linked to the use of thermostating procedures. Specifically, selecting the region where the thermostat is applied has a critical influence on the results. Whereas in the traditional homogeneous model the choices are limited to the fluid and/or the wall, in the inhomogeneous cases the reservoirs are also available, which permits to leave the region of interest, the channel, unperturbed.
\end{abstract}

\section{Introduction}
\label{sec-intro}

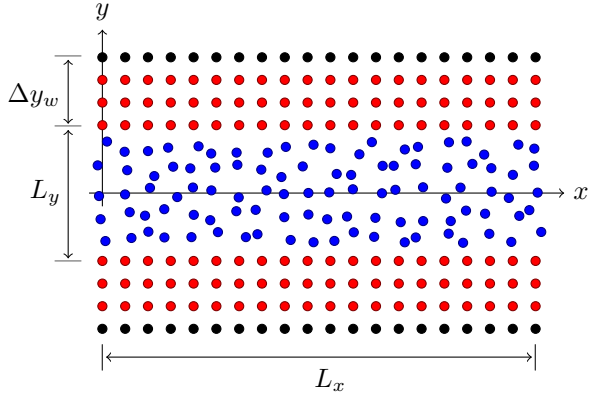
\begin{figure}[tbp]
\begin{center}
\begin{tikzpicture}[scale=0.3]
  \def\partsize{0.2}
 % \def\numparticles{400}

  % Coordinate system
  \node (X1) at (-1,5) {};
  \node (X2) at (21,5) {$x$};
  \node (Y1) at (0,4) {};
  \node (Y2) at (0,13) {$y$};
  \draw[->] (X1) to  (X2);
  \draw[->] (Y1) to  (Y2);

  \node (Ly1) at (-1.5,1.85) {---};
  \node (Ly2) at (-1.5,7.8) {---};
  \node (Ly1b) at (-1.5,1.7) {};
  \node (Ly2b) at (-1.5,8.2) {};    
  \node (Ly) at (-2.5,5) {$L_y$};
  \node (Lwall) at (-1.5,10.9) {---};
  \node (Lwallb) at (-1.5,7.65) {};
  \node (Lwallc) at (-1.5,11.3) {};  
  \node (Lwall) at (-3,9.35) {$\Delta y_w$};    
  \draw[<->] (Ly1b) to (Ly2b);    
  \draw[<->] (Lwallb) to  (Lwallc);    
    \node (Lx1) at (0,-2.25) {|};
    \node (Lx2) at (19,-2.25) {|};
    \node (Lx1b) at (-0.25,-2.25) {};
    \node (Lx2b) at (19.25,-2.25) {};    
    \node (Lx) at (10,-3.25) {$L_x$};
    \draw[<->] (Lx1b) to  (Lx2b);       

  \foreach \y in {-1}{% Lower wall
    \foreach \x in {0,...,19}{ 
      \fill [fill=black, draw=black] (\x,\y) circle(\partsize);
    }
  }
  \foreach \y in {0,...,2}{% Lower wall
    \foreach \x in {0,...,19}{ 
      \fill [fill=red, draw=red!50!black] (\x,\y) circle(\partsize);
    }
  }

 % It works but it does not look nice...
 % \foreach \i in {0,...,\numparticles}{
 %   \pgfmathsetmacro{\px}{rnd*20}
 %   \pgfmathsetmacro{\py}{rnd*5}
 %   \fill [fill=blue, draw=blue!50!black] ({\px},{2.5+\py}) circle(\partsize);
 % }
  \foreach \y in {3,...,7}{
    \foreach \x in {0,...,19}{ 
      \pgfmathsetmacro{\px}{rand*0.3}
      \pgfmathsetmacro{\py}{rand*0.3}
      \fill [fill=blue, draw=blue!50!black] ({\x+\px},{\y+\py}) circle(\partsize);
    }
  }

  \foreach \y in {8,...,10}{% Upper wall
    \foreach \x in {0,...,19}{ 
      \fill [fill=red, draw=red!50!black] (\x,\y) circle(\partsize);
    }
  }
  \foreach \y in {11}{% Upper wall
    \foreach \x in {0,...,19}{ 
      \fill [fill=black, draw=black] (\x,\y) circle(\partsize);
    }
  }

\end{tikzpicture}
\end{center}
\caption{Traditional molecular dynamics model of an homogeneous nanochannel flow. Fluid particles are shown in blue and solid particles in red and black.}
\label{fig-channel-homogeneous}
\end{figure}

The molecular dynamics (MD) configuration most commonly used to simulate nanochannel flows is shown in \Fig{channel-homogeneous}. The fluid particles are bounded by solid particles that model a wall, periodic boundary conditions are assumed in the streamwise and spanwise directions and the flow is driven applying a constant external force.  This driving mechanism has raised long-standing criticism for it requires a huge force to be applied, which generates an important amount of heat that, in turn, requires dissipative mechanisms (thermostating), and only represents an applied pressure difference when the pressure gradient is assumed to be constant everywhere \citep{Li1998}. Using this configuration implies assuming that the flow is streamwise homogeneous, which makes the problem easier by reducing it to one spatial dimension, the other two (streamwise and spanwise) being only statistical. On the other hand developing effects are eliminated from the beginning.

\tikzstyle{wall}=[circle,fill=red,draw=red!50!black] 
\tikzstyle{fluid}=[circle,fill=blue,draw=green] 
%\tikzstyle{vertexr}=[circle,red,draw=red,minimum size=20pt,inner sep=0pt] 
%\tikzstyle{element}=[circle,minimum size=6pt,inner sep=0pt]
%\tikzstyle{part}=[circle,blue,minimum size=6pt,inner sep=0pt]
%

Nevertheless, this configuration has been used for years to study the flow slip over solid surfaces and it is still widely used, see e.g. \cite{Detcheverry2013}. Molecular dynamics simulations are performed integrating the equations of motion of individual molecules. Introducing the interactions between them results in a system whose size is the number of molecules. Apart from these interactions, the external force driving the system is also introduced. As the system is isolated, the work performed by the external force driving the flow results in an increase of internal energy. Therefore, the only way to reach a steady state is through the introduction of a dissipative mechanism. This term is included assuming that the fluid is ``in contact with a thermal bath'' or ``a reservoir'' \citep{Thompson1997,Priezjev2010,Sendner2009,Kannam2013} which extracts energy from the system.
There are many possibilities, but the more commonly used in nonequilibrium MD simulations are the Langevin, Nosé-Hoover, Berendsen or DPD thermostats, see e.g. \cite{Allen1987} and \cite{Frenkel2001}. In the context of nanochannels the traditional approach \citep{Thompson1997} was to apply a thermostat in the whole channel while assuming the flow to be homogeneous in the streamwise and spanwise directions by applying periodic boundary conditions, as shown in \Fig{channel-homogeneous}. Wall particles are fixed but their interaction with fluid is kept, which results in fluid-solid interface friction.

%,Li2014
This model was improved including wall particles into the model, i.e. integrating their equations of motion too \citep{Liem1992,Khare1997,Kim2008,Martini,Bernardi2010,AlizadehPahlavan2011}. In this case, apart from the external force acting over all fluid particles in the streamwise direction, wall particles are also constrained to move around equilibrium positions by applying external forces to them (typically derived from quadratic potentials). This permits to apply a thermostat on the wall particles too. In an effort to minimize its impact, some authors \citep{AlizadehPahlavan2011} apply the thermostat to one solid layer only, the one being further from the fluid, shown in black color in \Fig{channel-homogeneous}. In fact, a rigorous derivation of this procedure has been developed in \cite{Kantorovich2008} and \cite{Kantorovich2008a} and named stochastic boundary conditions, showing that applying a thermostat on the external border of a solid accounts for the influence of an infinitely large solid thermal bath around it.

After this improvement, the next natural question is whether the fluid should be thermostats or not and there is a consensus in the literature about answering negatively \citep{Kannam2013}, considering that cooling through the walls is the only ``realistic'' \citep{Khare1997,Kim2008,Martini,AlizadehPahlavan2011,Yong2013b} dissipative mechanism ``mimicking real experiments'' \citep{Bernardi2010}. While in the case of the solid walls the thermal bath has a clear physical meaning, in the case of the fluid it has not. Besides, transport properties, viscosity and conductivity, shear stress and slip over a solid surface, measured by the slip length $L_s$, were shown to depend on the thermostat parameter $\Gamma$ \citep{Khare1997,Bernardi2010,AlizadehPahlavan2011}. Applying thermostats to shear flows has also been put into question by \cite{Padilla1996} because they remove heat at rates that are higher than the rate of conduction of heat across the fluid. The implication of this fact is the lack of time to maintain redistribution of energy across the system which implies that the steady states reached depend on the degrees of freedom the thermostat is coupled to. This effect is specially severe at high shear rates and the (perhaps over-pessimistic) conclusion by \cite{Padilla1996} is that the effort ``should be directed to simulate lower shear rates''.

At this point, two well-established facts collide: the application of an external force generates an important amount of heat and the application of a thermostat is unphysical. On top of that, assuming periodic boundary conditions, and therefore streamwise homogeneity, eliminates an intrinsic heat transfer mechanism present in any (nano)channel, the transfer of heat by convection, which makes the model definitively unrealistic.

Another important heat transfer mechanism is the generation by shear friction. While in macroscopic flows this term is usually negligible (except at very low Re numbers, i.e. creeping flows) at nanoscales this term is very important, specially at high shear rates. In that case the flow cannot be considered isothermal. 

To understand the balance of these mechanisms a simple model can be developed from the macroscopic energy conservation equation \citep{Batchelor1999}
\begin{equation}
\rho c_{p}\frac{dT}{dt}-\beta T \frac{dp}{dt} = -\nabla \cdot \mathbf{q} + \Phi 
\label{eq-temperature}
\end{equation}
where $c_{p}$ is the specific heat, $\rho$ the mass density, $\beta$ the (possibly temperature-dependent) thermal expansion coefficient, $p$ the pressure, $\mathbf{q}$ the internal heat flux, and $\Phi$ is the Rayleigh function representing the mechanical dissipation of energy in sheared motion, proportional to the viscosity and the square of velocity gradients in Newtonian fluids. It is worth noting that the second term of the left-hand side of \Eq{temperature} is only relevant in compressible fluids and it is negligible for nearly incompressible ones, representing a heat sink due to the energy required by dilatation to occur. It is also relevant that this equation is equivalent to the one obtained by a statistical treatment of molecular equations of motion \citep{Irving1950}, namely
\begin{equation}
\frac{\partial E}{\partial t} + \nabla \cdot \left( E \mathbf{u} + \mathbf{q} - \mathbf{u}\cdot \mathbf{\sigma} \right) = 0 
\label{eq-heat-balance}
\end{equation}
where $\mathbf{u}$ is the fluid velocity and $E=\rho(e+u^2/2+\psi)$ is the total energy per unit volume, that includes the internal energy $e$ and the external potential $\psi$ whose gradient is the applied driving force. Only after using kinetic and potential energy conservation \Eq{temperature} is obtained, which does not include the external force (whose work cancels with potential energy variation). The two terms on the left hand side of \Eq{temperature} come from the calculation of the internal energy variation, which, apart from the energy variation due to temperature changes, includes the energy variation by dilatation that vanishes for incompressible flows, as mentioned.

Assuming a one dimensional steady flow that is cooled (or heated) from the walls and modeling that by a Newton law with convection coefficient $h$, which accounts for the heat conduction in the fluid and the Kapitza resistance of the interface \citep{Barrat2002}, we get from \Eq{temperature}
\begin{equation} 
\dot{m} c_{p} \frac{ dT}{dx} - u A \beta T \frac{dp}{dx} = - P h (T-T_w) + \Phi A
\label{eq-1D-temperature}
\end{equation} 
where $\dot{m}$ is the mass flow rate across a section of the channel of length $L$, cross sectional area $A$ and perimeter $P$. The terms on the left-hand side represent convection heat transfer, the first term on the right-hand side cooling through walls and the second one viscous heating (given by $\Phi = \mu \gamma^2$ with $\mu$ the shear viscosity and $\gamma$ the shear rate). Once again, we emphasize that the second term on the left-hand side is negligible for incompressible flows but it turns out to play an important role otherwise, as it will be shown below.

The solution of \Eq{1D-temperature} can be obtained assuming the variables multiplying the temperature in the second term on the left-hand side to be constant, a restrictive hypothesis that, in any case, permits to understand the implications of neglecting it or not. Observe that in this case, equation \Eq{1D-temperature} has a uniform solution
\begin{equation} 
T_u = \frac{Ph}    { \left( Ph - u A \beta \frac{dp}{dx} \right)} T_w 
    + \frac{\Phi A}{ \left( Ph - u A \beta \frac{dp}{dx} \right)}.
\end{equation}
%{\color{blue} Then we can rewrite \Eq{1D-temperature} as 
% \begin{equation} 
% \dot{m} c_{p} \frac{ dT}{dx} =  - { \left( Ph - u A \beta \frac{dp}{dx} \right)} \left( T - T_u \right)
% \end{equation} 
% from where we obtain the solution.} 
In general when the flow enters the channel at a temperature $T_i$, it is cooled (or heated) along the channel, according to 
\begin{equation} 
\left( T - T_u \right) = \left( T_i - T_u \right) e^{-\frac{  { \left( Ph - u A \beta \frac{dp}{dx} \right)} L}{\dot{m} c_{p}} \frac{x}{L}},
\label{eq-temp}
\end{equation} 
reaching equilibrium asymptotically. 

This simple model permits to conclude that only if the inlet temperature is $T_u$ the flow can be homogeneous. Otherwise an exponential increase or decrease is to be expected. Besides, when the pressure gradient and the shear rate are negligible, i.e. $ - u A \beta \frac{dp}{dx} \ll Ph$ and $ \Phi A \ll P h$, taking the inlet temperature as the wall temperature results in an isothermal flow (the heat produced by shear is easily dissipated through the walls). 

This simple model also permits to understand how different configurations and flow driving mechanisms impact on the energy balance. In the traditional streamwise homogeneous model the terms in the left-hand side of \Eq{1D-temperature} vanish and the equilibrium between heat generation by shear and cooling through walls determines the fluid temperature. Applying a thermostat can be a way of representing the dissipative terms neglected (observe that the pressure gradient is negative and therefore the second term on the left-hand side of \Eq{1D-temperature} is dissipative).

%%%%%%%%%%%%%%%%%%%%%%%%%%%%%%%%%%%%%%%%%%%%%%%%%%%%%%%%%%%%%%%%%%%%%%%%%%%%%

In this article we discuss an alternative configuration and a driving mechanism similar to others recently proposed \citep{Nichollsa,Docherty2014,Ge2015}. In fact, many alternative driving mechanisms to study nanochannel flows have been proposed for years. The alternative proposed by \cite{Li1998} is the introduction of a ``reflecting particle membrane'', a Maxwell daemon that precludes (with a given probability) the particles to cross it in one direction, thus generating a pressure gradient. Whereas the method to drive the flow does not introduce additional energy into the system (thus not requiring thermostating), the pressure gradient is difficult to control (which is done through the given probability). Another approach \citep{Huang2006,Huang2008,Wang2012} is to fix the pressure in the external reservoirs by introducing rigid movable plates normal to the flow, which introduces a time dependent driving mechanism (the size of the reservoirs changes). In \cite{Zhang2009} the channel walls are moved as in the Couette problem whereas the flow is stopped by a cross sectional wall. 

The methods proposed by \cite{Nichollsa,Docherty2014,Ge2015} and the one we study here are small variations of the so-called ``reservoir method'' first proposed by \cite{Sun1992}. The pressure difference is generated applying a constant force in the reservoirs whereas they differ on how the temperature or density is controlled. Some authors introduce reservoirs while driving the flow with a constant force and a Nosé-Hoover thermostat applied in the whole domain, including the channel \citep{Su}.

However, these alternatives have not been widely used, specially in the study of hydrodynamic slip, one of the reasons being the important increase in the computational cost. Abandoning the streamwise periodicity makes the problem two-dimensional (the spanwise direction being still statistical). On the other hand, the advantages of these improved models have not been demonstrated. The case of an inlet temperature substantially higher than that of the walls was analyzed in \cite{Ge2015} but considering only thermal effects, the flow assumed to be hydrodynamically fully developed. On the other hand when the inlet temperature is similar to that of the walls, the fluid is heated inside the channel and hydrodynamic effects appear, e.g. the maximum velocity increases and the slip length decreases along the channel. We document these effects in \Sect{results} after detailed description of the methods in \Sect{methods}. We summarize the main conclusions in \Sect{conclusions}.

\section{Simulation method}
\label{sec-methods}

We construct a nanochannel flow by confining a monoatomic fluid between two smooth solid walls. Both the fluid and walls are composed by atoms which interact through the pairwise Lennard-Jones (LJ) potential,
\begin{equation} 
V_{ij}(r_{ij}) = \begin{cases} \: 4\epsilon_{ij} \left[ \left( \frac{\sigma_{ij}}{r_{ij}} \right) ^{12} - \left( \frac{\sigma_{ij}}{r_{ij}} \right) ^6 \right] \: , & \: r_{ij}<r_c \\
\: 0 \: , & \: r_{ij} \ge r_c \end{cases}
\label{eq-lj}
\end{equation} 
where $r_{ij} = \lvert \mathbf{r}_i - \mathbf{r}_j \rvert$ is the distance between atoms $i$ and $j$ whose positions are $\mathbf{r}_i$ and $\mathbf{r}_j$, and $\epsilon_{ij}$ and $\sigma_{ij}$ are the energy and length scales of the potential, respectively. The subscripts $i$ and $j$ indicate the atom types (hereinafter \textit{f} stands for fluid atoms and \textit{w} for wall ones), and the calculation of the interactions of each particle is truncated at a cut-off distance $r_c = 2.5 \: \sigma,$ since we have verified that the results do not change appreciably by increasing $r_c$. All the physical units in this work are expressed in LJ units (that is, in terms of the characteristic fluid length $\sigma = \sigma_{ff}$, energy $\epsilon = \epsilon_{ff}$, and atomic mass $m = m_f$). For liquid argon these values are $\sigma = 3.4$ $\unit{\AA }$, $\epsilon = 1.65 \times 10^{-21}$ J and $m = 6.63 \times 10^{-26}$ kg respectively.

In all our simulations, the interaction between wall and fluid atoms is chosen to be as intense as that between fluid monomers, $\epsilon_{fw} = \epsilon_{ff}$, which is considered highly hydrophilic \citep{Barrat1999,Priezjev2007,Yong2013}, and $\sigma_{fw} = \sigma_{ff}$. Each atom of the thermal wall is tethered around its equilibrium position via a quadratic potential,
\begin{equation}
V_{wall} \left( \mathbf{r} \right) = K_w \left( \mathbf{r} - \mathbf{r}_0 \right)^2 \: ,
\label{eq-spring}
\end{equation}
where $\mathbf{r}$ is the position of the wall atom and $\mathbf{r}_0$ its equilibrium position, and $K_w$ models the stiffness of the wall \citep{Asproulis2010}. For the current work we have used a value $K_w = 600 \: \epsilon / \sigma ^2$, which is inside the interval of values commonly used in MD studies, and has been proved to accomplish the two basic requirements for wall stiffness: (i) it is not too small, thus preventing the melting of the wall according to the Lindemann criterion \citep{Hansen1990}, and (ii) its associated frequency is low enough to allow the correct integration of the equations of motion of the wall atoms without reducing the time step \citep{Asproulis2010}. The mass of wall particles is $m_w = 10 \: m_f$ in order to reduce the vibration frequency, and they do not interact with each other to reduce computational time ($\epsilon_{ww} = 0$, since it does not affect significantly the wall dynamics).

Taking into account the volume accessible for the fluid, the average fluid mass density in all our simulations is $\rho_f = 0.86 \: m \sigma ^{-3}$. With regard to the walls, they form a face-centered cubic (fcc) lattice of number density equal to $3.90 \: \sigma^{-3}$, which implies an equilibrium nearest-neighbor distance of $0.71 \: \sigma$. The wall planes in contact with the fluid are (010) faces, with the [100] orientation of the fcc lattice aligned with the shear flow direction ($x$). The number of fluid, $N_f$, and wall atoms, $N_w$, vary from one studied configuration to another, and are detailed below.

All the simulations have been carried out using the LAMMPS package \citep{Plimpton1995}. The equations of motion are integrated using the velocity Verlet algorithm, with a time step of $\Delta t = 0.002 \: \tau$, where $\tau = \left( m\sigma^2/\epsilon \right)^{1/2}$ is the characteristic LJ time ($\tau = 2.16 \times 10^{-16}$ s for liquid argon). In the initial configuration the fluid particles are arranged in the positions of a fcc lattice, and the equilibration runs lasted typically $5 \times 10^5$ steps. Once the steady state is reached, a production run of a minimum of $10^6$ steps ($2\times 10^3 \: \tau$) is performed to average the data. The simulation domain is divided in bins of size $\Delta x = 1.5 \: \sigma$ and $\Delta y = 0.5 \: \sigma$ to discretize the collected data.

The components of the local stress tensor in each spatial bin have been computed following the Irving-Kirkwood method \citep{Irving1950}, that is,
\begin{align}
\mathbf{P} (\mathbf{r}_{bin}) & = \frac{1}{V_{bin}} \left< \sum\limits_{i\in bin}^{N_{bin}} m_i \left[ \mathbf{u}_i(t) - \mathbf{u}(\mathbf{r}_{bin},t) \right]\left[ \mathbf{u}_i(t) - \mathbf{u}(\mathbf{r}_{bin},t) \right]  \right> \notag \\
& + \frac{1}{2V_{bin}}  \left<  \sum\limits_{i\in bin}^{N_{bin}} \sum\limits_{j\ne i}^N \mathbf{r}_{ij}(t) \mathbf{F}_{ij}(t) \right>\label{eq-pressure_calculation_MD}
\end{align}
where the sum of the kinetic term includes the $N_{bin}$ particles which are inside the bin located at $\mathbf{r}_{bin}$ at time $t$, and the potential term involves the interaction of particles $i$ inside the bin with all the other atoms $j$ of the system (in or outside the bin); $\mathbf{F}_{ij}$ is the sum of internal forces exerted on $i$ by $j$, $\mathbf{u}(\mathbf{r}_{bin},t)$ the average velocity in the bin, and $V_{bin}$ its volume.

In order to control the temperature in some region of the computational domain the dissipative particle dynamics (DPD) thermostat is considered. This type of thermostat is considered to be particularly suitable for nonequilibrium MD \citep{Soddemann2003,Yong2013} since, among other advantages, it is a profile-unbiased thermostat \citep{Evans1986}; that is, does not need to assume a predetermined streaming velocity profile. This virtue of the DPD thermostat is due to the fact that it involves relative velocities between pairs of particles, $\mathbf{u}_{ij} = \mathbf{u}_i - \mathbf{u}_j$, instead of individual velocities as in other thermostats commonly used (e.g. Langevin thermostat). Hence, the equations of motion in the thermostated region are
\begin{equation} 
m_i \frac{d\mathbf{u}_i}{dt} \: = \: - \sum_{j \neq i} \mathbf{\nabla}_{\mathbf{r}_i} V_{ij} (r_{ij}) + \mathbf{F}^{D}_i + \mathbf{F}^{R}_i
\label{eq-dpd}
\end{equation} 
where two extra terms are added to the force resulting from the interatomic potential. $\mathbf{F}^{D}_i$ denotes the dissipative force on particle $i$ and $\mathbf{F}^{R}_i$ the corresponding random force. Both are expressed as a sum of pairwise contributions,
\begin{equation} 
\mathbf{F}^{D}_i = \sum_{j\neq i} \mathbf{F}^{D}_{ij} = - \sum_{j\neq i} \Gamma w^2(r_{ij})  \left( \hat{\mathbf{r}}_{ij} \cdot \mathbf{u}_{ij} \right) \hat{\mathbf{r}}_{ij}
\label{eq-dpd_dissipative}
\end{equation} 
\begin{equation} 
\mathbf{F}^{R}_i = \sum_{j\neq i} \mathbf{F}^{R}_{ij} = \sum_{j\neq i} \sqrt{2k_BT\Gamma} w(r_{ij}) \alpha_{ij} \hat{\mathbf{r}}_{ij}
\label{eq-dpd_random}
\end{equation} 
where $\hat{\mathbf{r}}_{ij}=\mathbf{r}_{ij}/|\mathbf{r}_{ij}|$, $\mathbf{r}_{ij} = \mathbf{r}_i - \mathbf{r}_j$, $T$ is the target temperature, $\Gamma$ the friction coefficient ($\Gamma = 1.0 \; m\tau^{-1}$ in our simulations), $\alpha_{ij}$ a Gaussian white noise variable that fulfills the condition $\alpha_{ij} = \alpha_{ji}$, and $w(r)$ is a weighting function of $r_{ij}$. The usual choice is 
\begin{equation} 
w(r_{ij}) = \begin{cases} \: 1- r/r_c \: , & \: r_{ij}<r_c \\
\: 0 \: , & \: r_{ij} \ge r_c \end{cases}
\label{eq-weight_function}
\end{equation} 

Previous features are common to all the models simulated in this work. In the rest of this section we describe the specificities of the various studied models, that differ essentially in the driving mechanism of the flow, the thermostated regions, and the geometry of the channel.

\subsection{A simple approach: the streamwise homogeneous (SH) flow model}
%\textbf{Model 1. Homogeneous flow}\\

The configuration geometry of the SH flow model is shown in \Fig{channel-homogeneous}. The channel length in the flow direction, $L_x$, varies from $200$ to $400 \: \sigma$ depending on the case, its width (measured as the distance between the wall planes in contact with the fluid) is $L_y = 30.0 \: \sigma$, and its depth $L_z = 10.0 \: \sigma$. Both the upper and lower walls consist of four fcc layers separated by a distance  $0.50 \: \sigma$ (that is, the wall thickness is $\Delta y_w = 1.50 \: \sigma$). Then, the number of fluid and wall atoms in the simulation cell vary from $N_f = 49980$ and $N_w = 39600$ (for $Lx\:=200 \sigma$) to $N_f = 99960$ and $N_w = 79200$ (for $Lx\:=400 \sigma$). Periodic boundary conditions are applied in $x$ and $z$ directions. As specified in \Fig{channel-homogeneous}, the plane $y=0$ cuts the channel through its center and $x=0$ at the entrance.

The flow is generated by applying a constant external force (per unit mass) $f_x$ in $x$ direction on all the fluid atoms. The interval of forces simulated in this work goes from $f_x = 0.010 \: \epsilon / m\sigma$ to $f_x = 0.040 \: \epsilon / m\sigma$. Modeling the walls as non-rigid allows for the heat generated by friction to be removed through them. The wall temperature is fixed to the value $T_w = 1.1 \: \epsilon/k_B$ by applying the DPD thermostat described above only to the wall atoms.

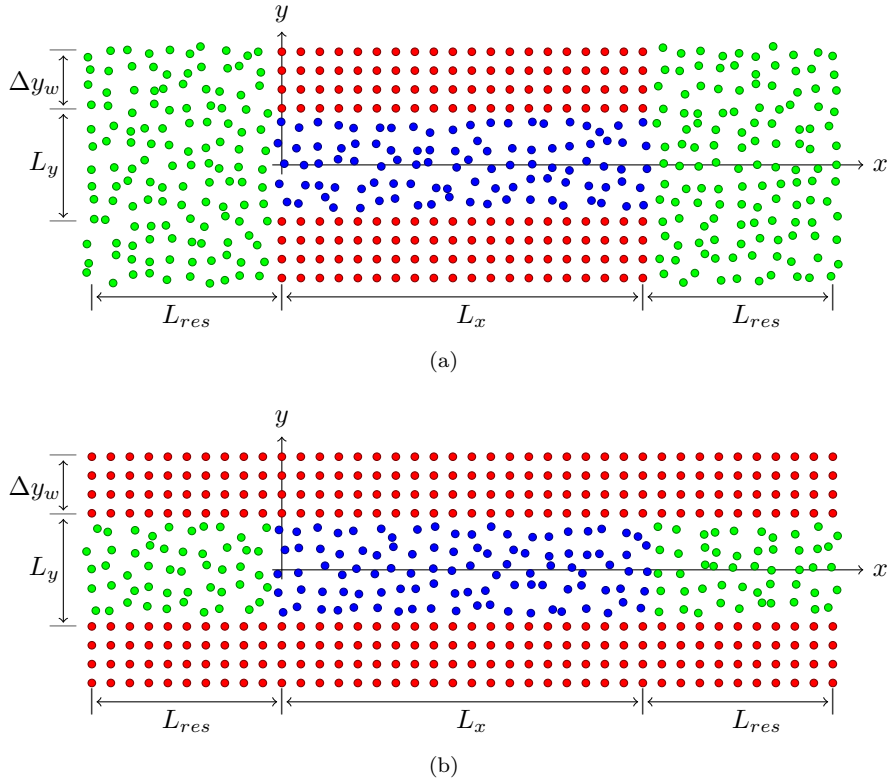
\begin{figure}[h!]
	\begin{center}
		\subfigure[] {
		\begin{tikzpicture}[scale=0.25]
		\def\partsize{0.2}
		% \def\numparticles{400}
		
		% Coordinate system
		\node (X1) at (9,5) {};
		\node (X2) at (41.5,5) {$x$};
		\node (Y1) at (10,4) {};
		\node (Y2) at (10,13) {$y$};
		\draw[->] (X1) to  (X2);
		\draw[->] (Y1) to  (Y2);
        \node (Ly1) at (-1.5,1.85) {---};
        \node (Ly2) at (-1.5,7.80) {---};
        \node (Ly1b) at (-1.5,1.70) {};
        \node (Ly2b) at (-1.5,8.20) {};    
        \node (Ly) at (-2.5,5) {$L_y$};
        \draw[<->] (Ly1b) to  (Ly2b); 
        \node (Lwall) at (-1.5,10.90) {---};
        \node (Lwallb) at (-1.5,7.65) {};
        \node (Lwallc) at (-1.5,11.30) {};  
        \node (Lwall) at (-3,9.35) {$\Delta y_w$};    
        \draw[<->] (Lwallb) to  (Lwallc);               
        \node (Lx1) at (10,-2.0) {|};
        \node (Lx2) at (29,-2.0) {|};
        \node (Lx1b) at (9.75,-2.0) {};
        \node (Lx2b) at (29.25,-2.0) {};    
        \node (Lx) at (20,-3.0) {$L_x$};
        \draw[<->] (Lx1b) to  (Lx2b);       
        \node (Lres1) at (0.0,-2.0) {|};        
        \node (Lres2) at (39.0,-2.0) {|};        
        \node (Lres1b) at (-0.25,-2.0) {};        
        \node (Lres1c) at (10.25,-2.0) {};        
        \node (Lres2b) at (28.75,-2.0) {};        
        \node (Lres2c) at (39.25,-2.0) {};                
        \node (Lres1) at (5,-3.0) {$L_{res}$};
        \node (Lres2) at (35,-3.0) {$L_{res}$};
        \draw[<->] (Lres1b) to  (Lres1c);       	
        \draw[<->] (Lres2b) to  (Lres2c);       	                			
		
		% \foreach \y in {-3,...,-1}{% Lower wall
		%   \foreach \x in {10,...,12}{ 
		%     \fill [fill=red, draw=red!50!black] (\x,\y) circle(\partsize);
		%   }
		% }
		\foreach \y in {-1,...,2}{%
			\foreach \x in {10,...,29}{ 
				\fill [fill=red, draw=red!50!black] (\x,\y) circle(\partsize);
			}
		}
		% \foreach \y in {-3,...,-1}{%
		%   \foreach \x in {27,...,29}{ 
		%     \fill [fill=red, draw=red!50!black] (\x,\y) circle(\partsize);
		%   }
		% }
		
		\foreach \y in {-1,...,11}{
			\foreach \x in {0,...,9}{ %Thermostated fluid
				\pgfmathsetmacro{\px}{rand*0.3}
				\pgfmathsetmacro{\py}{rand*0.3}
				\fill [fill=green, draw=green!50!black] ({\x+\px},{\y+\py}) circle(\partsize);
			}
		}
		
		\foreach \y in {3,...,7}{
			\foreach \x in {10,...,29}{ %Non thermostated fluid
				\pgfmathsetmacro{\px}{rand*0.3}
				\pgfmathsetmacro{\py}{rand*0.3}
				\fill [fill=blue, draw=blue!50!black] ({\x+\px},{\y+\py}) circle(\partsize);
			}
		}
		
		\foreach \y in {-1,...,11}{
			\foreach \x in {30,...,39}{  %Thermostated fluid
				\pgfmathsetmacro{\px}{rand*0.3}
				\pgfmathsetmacro{\py}{rand*0.3}
				\fill [fill=green, draw=green!50!black] ({\x+\px},{\y+\py}) circle(\partsize);
			}
		}

		% \foreach \y in {11,...,13}{% Upper wall
		%   \foreach \x in {10,...,12}{ 
		%     \fill [fill=red, draw=red!50!black] (\x,\y) circle(\partsize);
		%   }
		% }
		\foreach \y in {8,...,11}{%
			\foreach \x in {10,...,29}{ 
				\fill [fill=red, draw=red!50!black] (\x,\y) circle(\partsize);
			}
		}
		% \foreach \y in {11,...,13}{%
		%   \foreach \x in {27,...,29}{ 
		%     \fill [fill=red, draw=red!50!black] (\x,\y) circle(\partsize);
		%   }
		% }
		
		\end{tikzpicture}
			\label{fig-channel-developing}}
	\subfigure[]{
		\begin{tikzpicture}[scale=0.25]
		\def\partsize{0.2}
		% \def\numparticles{400}
		
		% Coordinate system
		\node (X1) at (9,5) {};
		\node (X2) at (41.5,5) {$x$};
		\node (Y1) at (10,4) {};
		\node (Y2) at (10,13) {$y$};
		\draw[->] (X1) to  (X2);
		\draw[->] (Y1) to  (Y2);
		\node (Ly1) at (-1.5,1.85) {---};
		\node (Ly2) at (-1.5,7.80) {---};
		\node (Ly1b) at (-1.5,1.7) {};
		\node (Ly2b) at (-1.5,8.20) {};    
		\node (Ly) at (-2.5,5) {$L_y$};
		\draw[<->] (Ly1b) to  (Ly2b);    
		\node (Lwall) at (-1.5,10.90) {---};
		\node (Lwallb) at (-1.5,7.65) {};
		\node (Lwallc) at (-1.5,11.30) {};  
		\node (Lwall) at (-3,9.35) {$\Delta y_w$};    
		\draw[<->] (Lwallb) to  (Lwallc);    		
		\node (Lx1) at (10,-2.0) {|};
		\node (Lx2) at (29,-2.0) {|};
		\node (Lx1b) at (9.75,-2.0) {};
		\node (Lx2b) at (29.25,-2.0) {};    
		\node (Lx) at (20,-3.0) {$L_x$};
		\draw[<->] (Lx1b) to  (Lx2b);       
		\node (Lres1) at (0.0,-2.0) {|};        
		\node (Lres2) at (39.0,-2.0) {|};        
		\node (Lres1b) at (-0.25,-2.0) {};        
		\node (Lres1c) at (10.25,-2.0) {};        
		\node (Lres2b) at (28.75,-2.0) {};        
		\node (Lres2c) at (39.25,-2.0) {};                
		\node (Lres1) at (5,-3.0) {$L_{res}$};
		\node (Lres2) at (35,-3.0) {$L_{res}$};
		\draw[<->] (Lres1b) to  (Lres1c);       	
		\draw[<->] (Lres2b) to  (Lres2c);       	
		
		\foreach \y in {-1,...,2}{% Lower wall
			\foreach \x in {0,...,39}{ 
				\fill [fill=red, draw=red!50!black] (\x,\y) circle(\partsize);
			}
		}
		
		\foreach \y in {3,...,7}{
			\foreach \x in {0,...,9}{ %Thermostated fluid
				\pgfmathsetmacro{\px}{rand*0.3}
				\pgfmathsetmacro{\py}{rand*0.3}
				\fill [fill=green, draw=green!50!black] ({\x+\px},{\y+\py}) circle(\partsize);
			}
			\foreach \x in {10,...,29}{ %Non thermostated fluid
				\pgfmathsetmacro{\px}{rand*0.3}
				\pgfmathsetmacro{\py}{rand*0.3}
				\fill [fill=blue, draw=blue!50!black] ({\x+\px},{\y+\py}) circle(\partsize);
			}
			\foreach \x in {30,...,39}{  %Thermostated fluid
				\pgfmathsetmacro{\px}{rand*0.3}
				\pgfmathsetmacro{\py}{rand*0.3}
				\fill [fill=green, draw=green!50!black] ({\x+\px},{\y+\py}) circle(\partsize);
			}
		}
		
		\foreach \y in {8,...,11}{% Upper wall
			\foreach \x in {0,...,39}{ 
				\fill [fill=red, draw=red!50!black] (\x,\y) circle(\partsize);
			}
		}
		
		\end{tikzpicture}
			\label{fig-channel-developing2}}
	\end{center}
	\caption{(a) New configuration suggested in this work for inhomogeneous MD simulations of nanochannel flows. Fluid particles in the reservoirs (in green) are thermostated and allowed to move freely in the $y$ direction. The properties of the fluid inside the channel (atoms in blue) varies along the $x$ direction. Solid particles are shown in red. (b) Alternative configuration where vertical motion is constrained in the left and right reservoirs by extensions of the channel walls.}	
\end{figure}

\subsection{Abandoning homogeneity: the streamwise inhomogeneous force driven (SIFD) flow model}

In a first step towards a more realistic model, the homogeneity hypothesis is abandoned and the configuration shown in \Fig{channel-developing} is studied. 
We consider a central channel of the same dimensions as in the SH model, $ L_x \times L_y \times L_z = 200-400 \: \sigma \times 30 \: \sigma \times 10 \: \sigma$, limited by the same fcc walls of thickness $\Delta y_w = 1.50 \: \sigma$. This is the domain of interest, where the fluid properties are extracted. But now we add two open reservoirs of length $L_{res} = 50 \: \sigma$ outside it, both on the left and on the right of the channel, where the fluid can move freely in the vertical ($y$) direction. Periodic boundary conditions are applied in the three directions. Again, we choose the origin in such a way that $y=0$ at the center of the channel and $x=0$ at the entrance (and, then, $x$ coordinates take negative values at the left reservoir).

We apply a DPD thermostat, \Eqs{dpd}{weight_function}, to the fluid particles, but only when they are in the reservoirs outside the domain of interest. In such a way, we fix the temperature of the fluid at the inlet to be $T_{in} = 1.1 \: \epsilon /k_B$ and we leave the fluid completely free inside the channel. As in SH model, the walls are also thermostated to  $T_w = 1.1 \: \epsilon/k_B$, and the flow is driven by a constant external force $f_x$ exerted on every fluid atom in the whole simulated domain.

Unlike the SH flow model, the SIFD model allows for the evolution of the fluid properties along the channel, like the local temperature, and therefore incorporates the heat transfer by convection. The presence of the reservoirs also makes it possible to analyze the channel entrance effects, like pressure losses. The basic idea behind a model like this is to explicitly separate the domain of interest, where the system evolves according its natural dynamics, without being restricted by artificial forces or constraints, from the surroundings where constraints are applied to induce the desired fluid conditions at the entrance of the region of interest. This is, precisely, the great difficulty when periodicity is abandoned in MD: how to impose the proper boundary conditions to couple both regions adequately. In fact, this is also the main challenge to build hybrid models which couple molecular dynamics with continuum dynamics \citep{Hadjiconstantinou1997,Werder2005,Weinan2007,Mohamed2010}.

As mentioned in \Sect{intro}, other works have previously proposed different boundary conditions for generating inhomogeneous flows on nanopores and one of the reasons why the use of this kind of models is not generalized is that enlarging a system to include a region outside the domain of interest has a computational cost. In our case, the number of fluid atoms grows to $N_f = 78300$ (for $L_x = 200 \: \sigma$) and $N_f = 128280$ (for $L_x = 400 \: \sigma$). Nevertheless, it is affordable given the tremendous amount of computing power available in supercomputers and the maturity of the simulation software.

A simplification of the configuration of this SIFD model is presented in \Fig{channel-developing2}. Again we consider a channel of length $L_x$ and we add left and right reservoirs that are extensions of the channel. In these reservoirs the flow is thermostated and periodic boundary conditions are applied but the vertical motion is constrained by (fictitious) extensions of the channel walls. This configuration does not account for hydrodynamic entrance effects but it will be important to understand the relation to the SH model.

There are therefore two types of SIFD models according to the type of reservoir: the SIFD model with open reservoirs (SIFD-OR) and with closed reservoirs (SIFD-CR).

\subsection{Leaving the channel unperturbed: the streamwise inhomogeneous pressure driven (SIPD) flow model}

Finally, we have simulated another model in which body forces no longer exist inside the channel. The configuration is the same shown in \Fig{channel-developing}, but now a pressure gradient is generated between the inlet and outlet reservoirs by applying a force only on those fluid atoms located far from the channel. In particular, we apply an external force of magnitude
\begin{equation} 
f_x \: = \: \frac{\Delta p}{\rho_f \: \Delta z}
\label{eq-force_model3}
\end{equation} 
on all the fluid atoms located inside two regions of width $L_{res}/3$ at the edges of the simulation cell (one at the beginning of the left reservoir and the other at the end of the right one), but not outside them. $\Delta z = 2L_{res}/3$ is the total length of both regions where the force is applied, and $\Delta p$ is the pressure difference created. The DPD thermostat described above is applied in the reservoirs of length $L_{res}$ to fix the temperature to $1.1 \: \epsilon/k_B$. In this configuration the heat generated by the external force is dissipated inplace, as far from the channel as possible trying to minimize the disturbance caused in the system.
 
\section{Results}
\label{sec-results}

\subsection{Homogeneous flow}

\begin{figure}[h!]
	\begin{center}
	\includegraphics[width=0.7\textwidth]{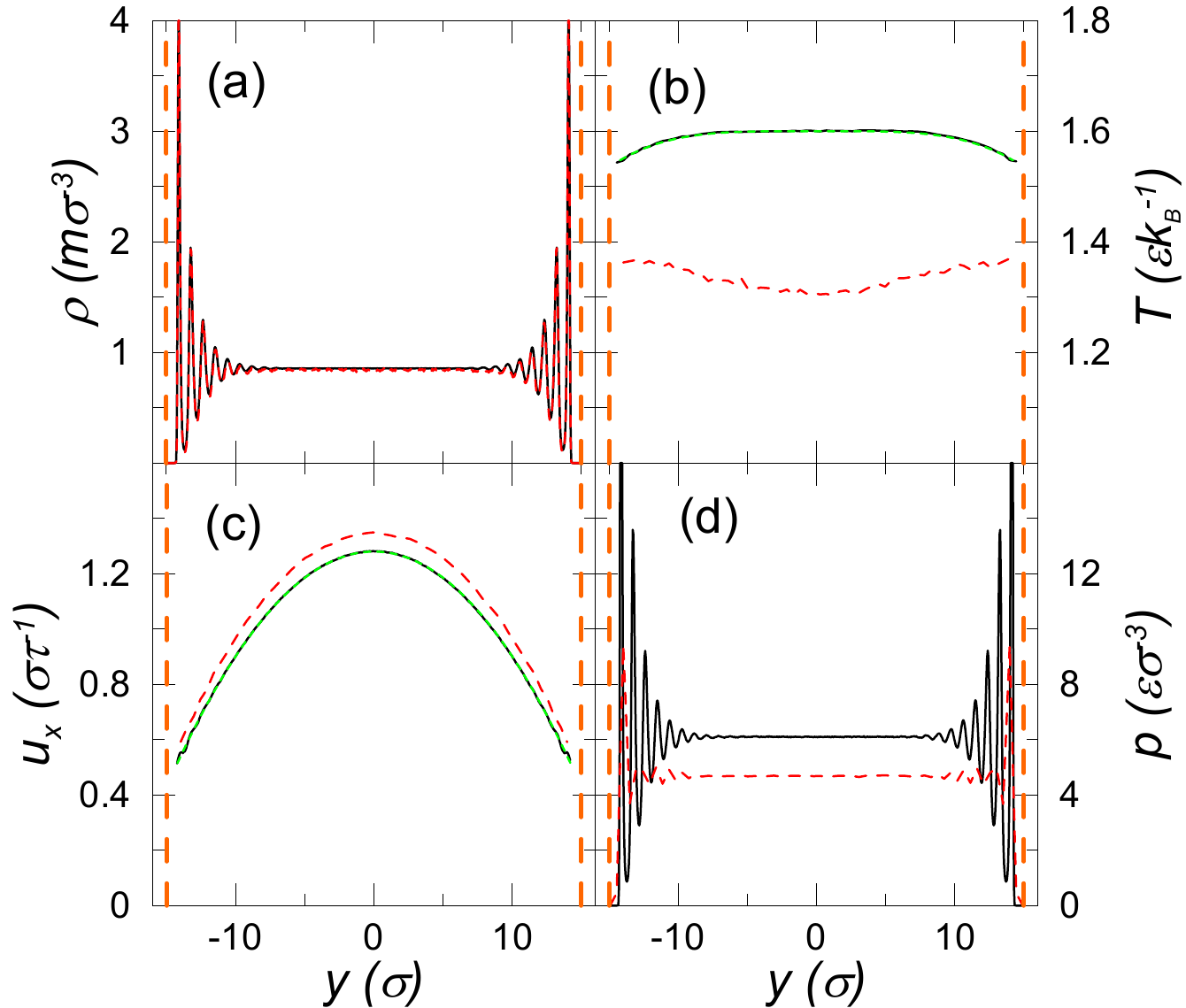}
	\caption{Averaged fluid profiles in the section $x=150$ across the flow direction, obtained with the SH (black solid line) and the SIFD-CR model (red dashed line), for an applied force $f_x = 0.020 \: \epsilon/m\sigma$ and $L_x = 200 \: \sigma$: (a) fluid density, (b) temperature, (c) streaming velocity, and (d) pressure. The dotted green curves in panels (b) and (c) are the fits of the homogeneous temperature \Eq{temperature_quartic}) and velocity \Eq{velocity_quadratic}) profiles, respectively (see text). Vertical dashed lines indicate the position of the innermost walls layers.}
	\label{fig-homo_vs_inhomo_y}
	\end{center}	
\end{figure}

The fluid properties obtained by atomistic simulations of periodic homogeneous flows in nanochannels are rather well understood. In \Fig{homo_vs_inhomo_y} we show the averaged density, temperature, velocity, and pressure profiles for the SH model. As previously mentioned, this model maintains the wall temperature fixed but it does not thermostat the fluid, which is widely accepted to be the more realistic option for homogeneous flows \citep{Khare1997,Kim2008,Martini,AlizadehPahlavan2011,Kannam2013}. Nevertheless, the evacuation of the viscous heat through the walls does not avoid a significant temperature increase in the fluid, as we shall see shortly. The constant force applied to induce the flow in this case has been $f_x = 0.020 \: \epsilon/m\sigma$, which despite being a value in the range of those commonly used in MD exceeds the gravity force by a factor of $1.5 \times 10^{11}$.

The fluid density shows a clear layered structure (with at least six marked layers separated by a distance $\sim 0.9 \: \sigma$) in the region near to the atomic walls, where the surface effects are visible. On the other hand, it is constant and equals the bulk value in the center of the channel. This fact suggests that the channel is wide enough to assume the continuum equations to be valid at this scale \citep{Bocquet2010}. In particular, the streaming velocity can be determined from the momentum equation 

\begin{equation} 
\rho u_x \frac{\partial u_x}{\partial x} \: = \: - \frac{\partial p}{\partial x} + \mu \frac{\partial^2 u_x}{\partial y^2} + \rho f_x
\label{eq-momentum}
\end{equation} 
where $\rho$ is the fluid mass density, and $u_x$ the $x$ component of the streaming velocity. Since there is no variation along the channel, the well-known quadratic profile is recovered,
\begin{equation} 
u_x \left( y \right)  \: = \: \frac{\rho f_x}{2\mu} \: \left( \frac{h^2}{4} - y^2 + hL_s \right) 
\label{eq-velocity_quadratic}
\end{equation} 
where $h$ is the distance between the solid-liquid interfaces at the top and bottom walls. The position of the solid-liquid interface (that is, the point of closest approach where the boundary condition is imposed) is not well defined. To take into account the excluded volume effects, we locate the interface at a distance of $0.5 \: \sigma$ from the wall innermost fcc planes \citep{Priezjev2007}, and then $h = 29 \: \sigma$. The slip length $L_s$ is defined as the additional length, relative to the interface, at which the linearly extrapolated fluid tangential velocity vanishes,
\begin{equation}
\left\vert \frac{\partial u_x}{\partial y} \left( y= \pm h/2 \right) \right\vert \: L_s \: = \: u_s
\label{eq-slip_length}
\end{equation}
with $u_s = u_x(\pm h/2)$ the slip velocity at the interface. As it can be seen in \Fig{homo_vs_inhomo_y}, the solution in \Eq{velocity_quadratic} fits accurately the velocity profile assuming a value around $\mu \sim 2.4 \: \epsilon \tau \sigma^{-3}$ for the viscosity, which coincides with that obtained from the simulated shear stress, $\mu \: = \: P_{xy} \left( \partial u_x / \partial y \right)^{-1} = 2.45 \pm 0.10 \: \epsilon \tau \sigma^{-3}$, and is close to those obtained with similar models \citep{Thompson1997}. The simulated flow rate is then consistent with that obtained from the quadratic profile
\begin{equation}
Q \: = \: \frac{L_z h^3}{12\mu} \rho f_x \: \left( 1 + \frac{6L_s}{h} \right)
\label{eq-flow_rate}
\end{equation}\

With regard to the temperature, again the homogeneity simplifies the energy balance equation,
\begin{equation} 
\rho c_p u_x \frac{\partial T}{\partial x} - \beta T u_x \frac{\partial p}{\partial x} \: = \: \mu \left( \frac{\partial u_x}{\partial y} \right)^2 + \kappa \left( \frac{\partial^2 T}{\partial y^2} \right) ,
\label{eq-energy}
\end{equation} 
which reduces to a quartic profile for the temperature \citep{Todd1997},
\begin{equation} 
T \left( y \right)  \: = \: \frac{\rho^2 f_x^2}{12\kappa \mu} \: \left[ \left( \frac{h}{2} \right)^4 - y^4 +\frac{h^3 L_K}{2} \right]
\label{eq-temperature_quartic}
\end{equation} 
with $\kappa$ the thermal conductivity of the system and $L_K$ the Kapitza length, which is defined equivalently to the slip length in \Eq{slip_length} but changing $u_x$ by $T$ \citep{Khare2006}. As it can be seen in \Fig{homo_vs_inhomo_y} the temperature profile is satisfactorily fitted by a quartic function.

\subsection{Streamwise inhomogeneous force-driven flow}

\begin{figure}[h!]		
	\begin{center}
		\includegraphics[width=0.7\textwidth]{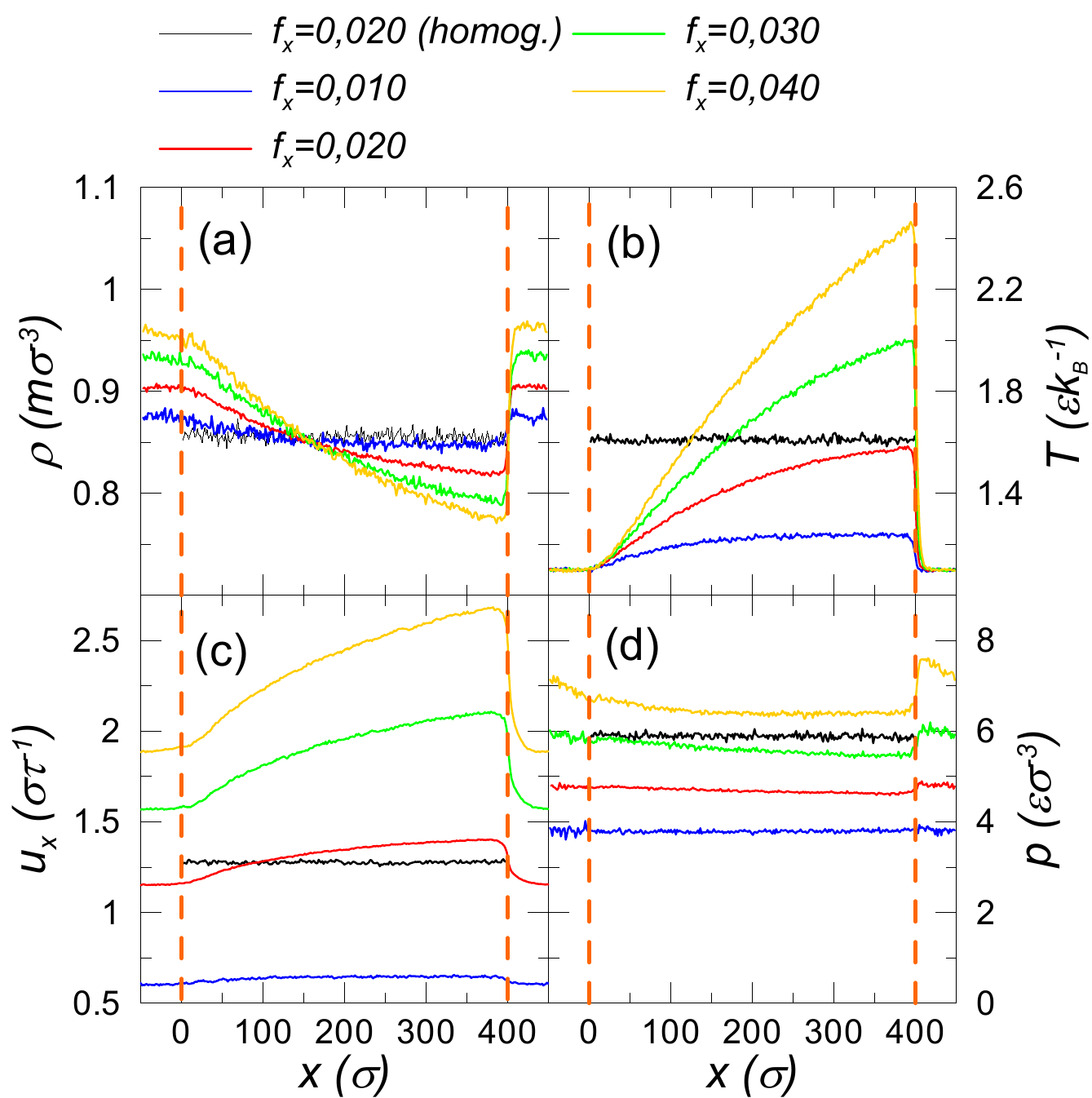}
		\caption{Averaged fluid profiles in the plane $y=0$ along the flow direction, obtained with the SH (black line) and the SIFD-CR model for different applied force values: (a) fluid density, (b) temperature, (c) streaming velocity, and (d) pressure. Vertical dashed lines indicate the entrance and exit of the channel.}
		\label{fig-homo_vs_inhomo_x}
	\end{center}	
\end{figure}

Unless stated explicitly, in this subsection we discuss the results obtained using the SIFD-CR model (that is, with the outer fixed-temperature reservoirs confined by the walls, the configuration shown in \Fig{channel-developing2}). We will show below that the results with the SIFD-OR model (using the open-reservoirs configuration in \Fig{channel-developing}) are qualitatively similar, and will discuss the slight differences. From \Fig{homo_vs_inhomo_y}, it could seem that the differences between homogeneous and non-homogeneous models are not that noticeable (except for the temperature). But we must take into account that, whereas the homogeneous profiles remain unaltered along the channel, the fluid properties in the inhomogeneous model evolve through $x$. In \Fig{homo_vs_inhomo_y}, then, we present the fluid profiles in a particular section ($x=150 \: \sigma$, far enough from the entrance) only as an example. It is more convenient to analyze the results along the direction of the flow, as we do in \Fig{homo_vs_inhomo_x}.

One of the main distinctive features of inhomogeneous models is their compressibility: the fluid density $\rho$ diminishes significantly along the channel, and this reduction is more pronounced for higher external forces, as expected. Note that only the values inside the non-thermostated channel have physical meaning; those in the reservoirs are artificial because of the applied thermostat and the imposed periodicity. On the contrary, the pressure is almost constant in the flow direction (a slight gradient is observed only for very high forces). The reason of this behavior seems to be in the configuration used: since both reservoirs are limited by walls, viscous forces are high enough to equilibrate the external force in these regions (last two terms in \Eq{momentum}), and thus an appreciable pressure difference is not created between channel ends. On the contrary, we will see that a clear pressure gradient arises when $f_x$ is applied in open reservoirs, in which friction is much less important, as it is the case in \Fig{channel-developing}. 

\begin{figure}[htbp]
	\begin{center}
		\subfigure[]{
			\includegraphics[width=0.45\textwidth]{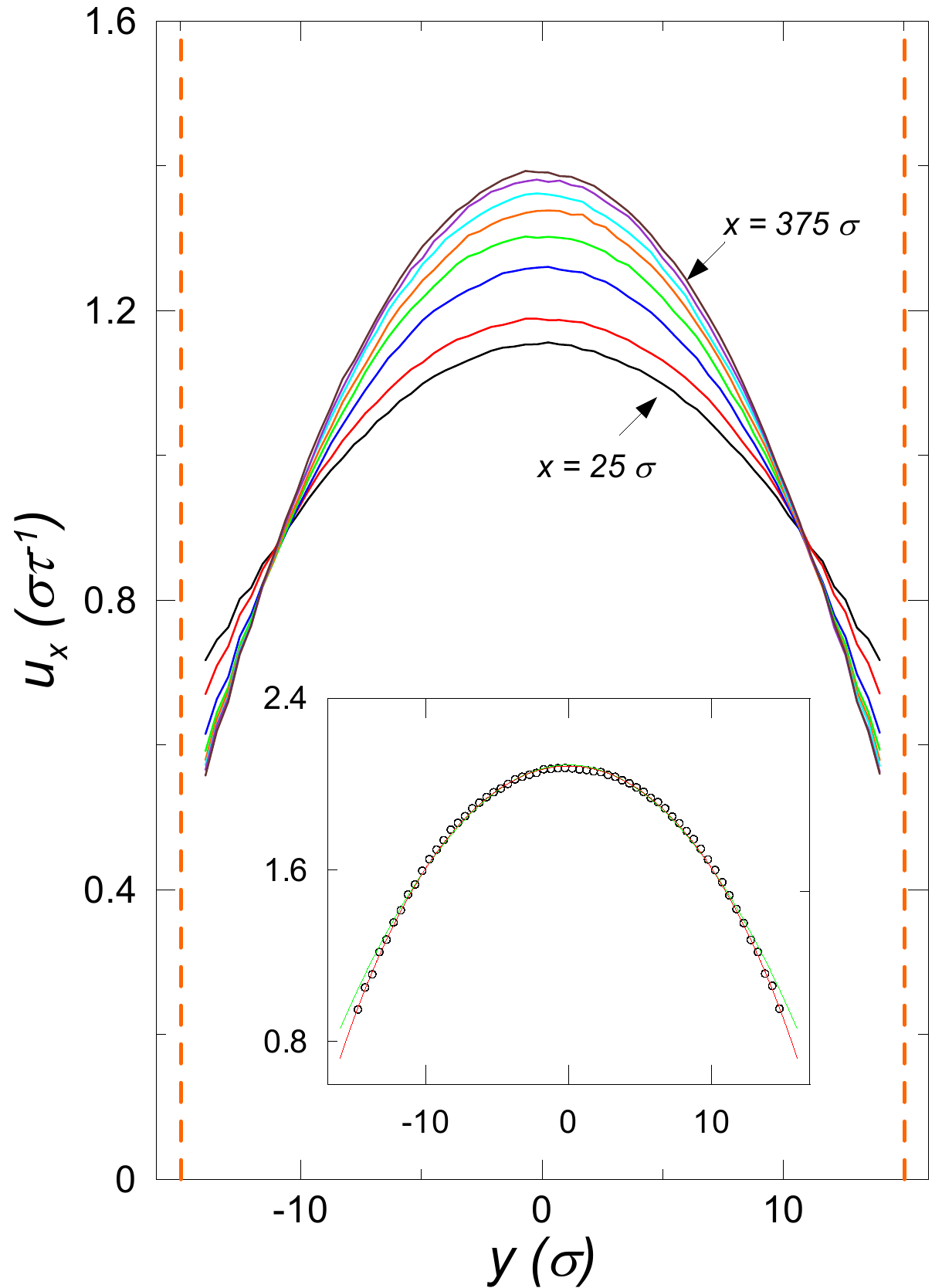}
			\label{fig-velocity_developing}}
		\subfigure[]{
			\includegraphics[width=0.45\textwidth]{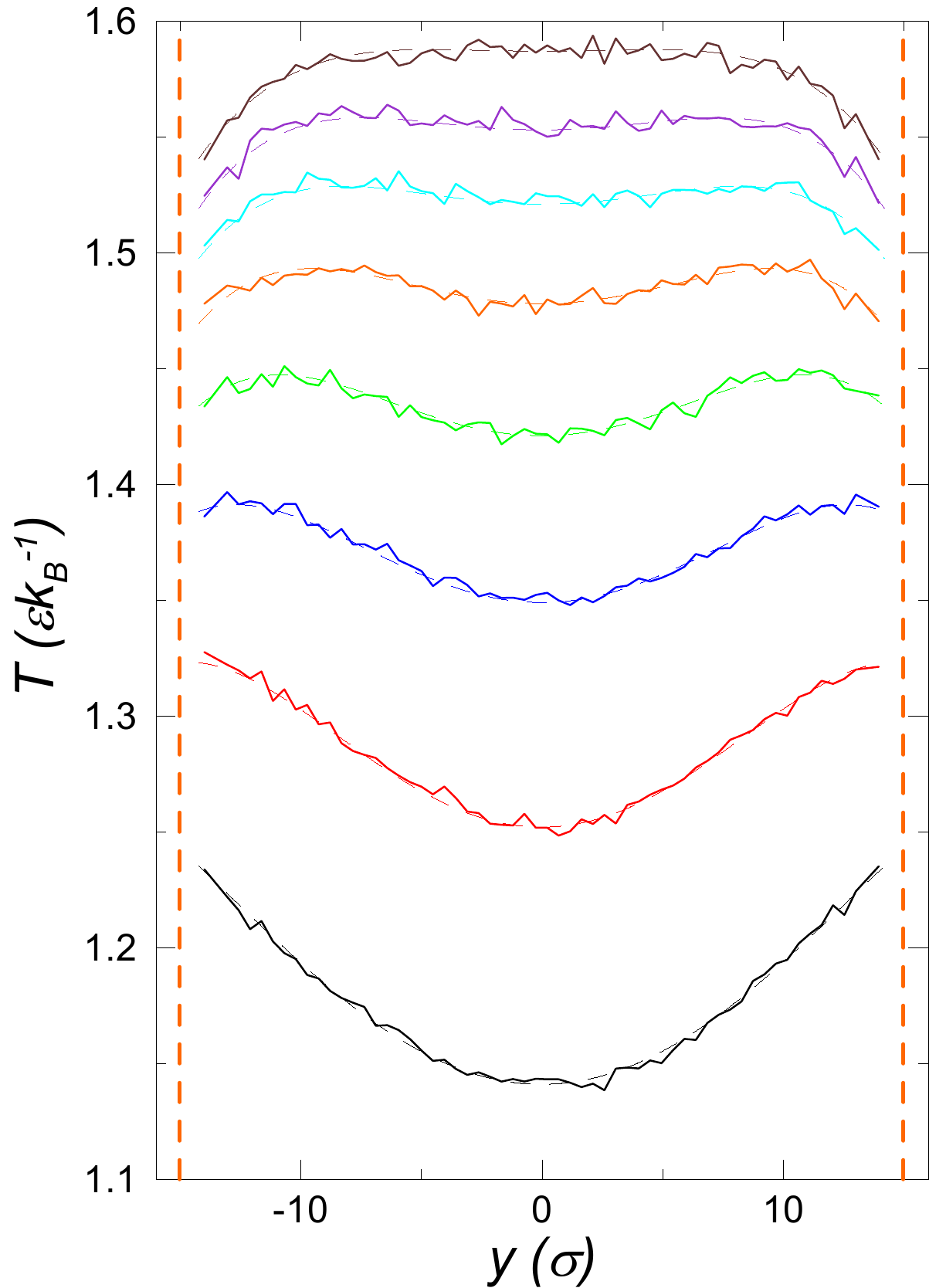}
			\label{fig-temperature_developing}}
		\caption{(a) Velocity and (b) temperature profiles across different channel sections obtained using the SIFD-CR model with a force $f_x=0.02 \: \epsilon / m\sigma$ in a channel of length $L_x=400 \: \sigma$. The eight curves corresponds to sections from $x=25 \: \sigma$ to $x=375 \: \sigma$ (from bottom to top, in successive steps of $50 \: \sigma$). Inset in (a) shows the velocity profile in $x=75 \: \sigma$ for a force $f_x=0.04 \: \epsilon / m\sigma$ in a channel of length $L_x=400 \: \sigma$ (empty circles), together with the best quadratic (\Eq{velocity_quadratic}, green curve) and hyperbolic (\Eq{velocity_hypcosin}, red curve) fits. Dashed curves in (b) are the fits of temperature profiles with the form in \Eq{temperature_quartic+quadratic}.}
	\end{center}	
\end{figure}

The compressibility of the flow allows the variation of the velocity along the channel, in such a way that the mass flow rate is constant. As it is shown in \Fig{homo_vs_inhomo_x} (c) and \Fig{velocity_developing}, when the fluid enters the non-thermostated channel its velocity profile starts to develop (in the reservoirs, the thermostat restrains the fluid and its velocity remains constant). Due to the friction, velocity gradually reduces in the regions near the walls, and therefore the fluid is accelerated at the center of the section to maintain the mass flow rate (\Fig{velocity_developing}). As a consequence the slip decreases (and shear rate $\gamma$ increases) in the streamwise direction. The shear continues to grow downstream until the friction force equilibrates the external force; downstream this entry region, the flow is fully developed. Whether a given channel is long enough to consider the flow hydrodynamically developed should be determined when designing the simulation of nanoscale flows. On the basis of the results of this work, forces higher than $0.02 \: \epsilon/m\sigma$ requires entry lengths longer than $400 \: \sigma$. However, it is common to find considerably shorter channels in the literature in which entrance effects are neglected.

Another noticeable impact of the change in the configuration is that the well-established quadratic profile for the velocity in \Eq{velocity_quadratic} may no longer be valid for non-homogeneous flows, since the first term in momentum equation, \Eq{momentum}, does not vanish. The solution becomes significantly more complex, but as a first approximation one can assume that the velocity gradient $\partial u_x / \partial x$ does not depend on $y$ (this is almost exactly true in our simulations). In this case, the new solution has the form
\begin{equation} 
u_x \left( y \right)  \: = u_0 \left[  1 - A \; {\rm cosh} \left( \lambda_0  y \right) \right]
\label{eq-velocity_hypcosin}
\end{equation}
where
\begin{equation*} 
u_0 = \frac{f_x}{\partial u_x / \partial x} \qquad \text{,} \qquad  \lambda_0^2 = \frac{\rho \partial u_x / \partial x}{\mu}
\end{equation*} 
and
\begin{equation*} 
A^{-1} = {\rm cosh} \left( \frac{\lambda_0 h}{2} \right) + L_s \lambda_0 {\rm sinh} \left( \frac{\lambda_0 h}{2} \right)
\end{equation*} 
This solution reduces to \Eq{velocity_quadratic} when the velocity gradient is small. Although we have indeed confirmed that the hyperbolic profile fits better the results than the quadratic one for high forces, the difference in our simulations is small (it is only noticeable near the boundary; see the inset in \Fig{velocity_developing}). Nevertheless, it should be taken into account in future studies or for more intense driving forces, as an accurate velocity fit can affect the calculation of the slip length. Consequently, the volumetric flow rate expression in \Eq{flow_rate}, used regularly in the literature for obtaining the slip length from experimental flow rate measures \citep{Li2014}, should be also modified to take into account the non-homegeneity, to be
\begin{equation}
Q \: = \: L_z u_0 \left[ h - \frac{2A}{\lambda_0} \; {\rm sinh} \left( \lambda_0  \frac{h}{2} \right) \right]
\label{eq-flow_rate_hypsin}
\end{equation}
More important than the change in the expression is the fact that the flow rate is variable in the streamwise direction.

But over all the features of this model for inhomogeneous flows, there is one that makes it clearly more realistic than the traditional homogeneous models: it incorporates the fluid cooling by convection along the channel. As it occurs in real (nano)channels, the fluid at the entrance is colder than at the outlet, and this makes the temperature to gradually rise due to the viscous heat generated by friction. The evolution of $T$ in \Fig{homo_vs_inhomo_x}(b) is qualitatively similar to the simple unidimensional model drafted in \Sect{intro}, \Eqs{1D-temperature}{temp}, and tends asymptotically to a constant value. Again, we have found that, for high external forces, the length of full thermal development is longer than the simulated channels. It must be noted that the asymptotic value to which $T$ tends in the case of $f_x = 0.020 \: \epsilon / m\sigma$ coincides with the temperature of the homogeneous model with the same applied force. This fact leads us to conclude that, while representing convection by a thermostat when assuming an homogeneous channel is a crude approximation, modeling cooling only through walls also fails to describe heat transfer, especially at the entrance, and overestimates the temperature at the channel. Only in low-shear regime (forces lower than $f_x = 0.010 \: \epsilon / m\sigma$ in our model) temperature is approximately uniform and the role of convection less important, as thermal conduction is effective enough, a case in which the SH model without fluid thermostating provides similar results.
 
It is also interesting to analyze how the temperature distribution across the channel varies with $x$. Results presented in \Fig{temperature_developing} show marked differences between the profiles as the flow progresses. At the inlet, where the convective cooling is intense, the thermal jump at the boundary is very pronounced and heat is transferred by conduction from the walls to the center of the channel. Only at sections where convection ceases to play a major role the temperature profile resembles that obtained in homogeneous models (see \Fig{homo_vs_inhomo_y}(b)). Understanding this behavior requires noticing that the convective term in the energy \Eq{energy} does not vanish now. Only to find an approximate solution, we can assume that specific heat, density, viscosity and thermal conductivity do not vary appreciably with $y$; that the temperature gradient is also approximately independent on $y$, and the pressure gradient negligible (the last two hypothesis have been checked to be valid here). Finally, for the sake of simplicity we take the solution \Eq{velocity_quadratic} for the velocity (since we have seen that solution \Eq{velocity_hypcosin} offers similar results for the simulations presented in this work). With these approximations, we get  
\begin{equation} 
T \left( y \right)  \: = \: a_2 y^2 \: - \: a_4 y^4 \: + \: a_0 
\label{eq-temperature_quartic+quadratic}
\end{equation} 
where 
\begin{equation*} 
a_2 \: = \: \frac{f_x}{4 \mu} \frac{\rho^2}{\kappa} c_p \frac{\partial T}{\partial x} \left( \frac{h^2}{4} + hL_s \right) \qquad \text{,} \qquad
a_4  \: = \: \frac{\rho f_x}{24 \mu} \: \left( \frac{\rho}{\kappa} c_p \frac{\partial T}{\partial x} + 2 \frac{\rho f_x}{\kappa} \right) \quad
\end{equation*} 
and 
\begin{align*} 
  a_0  \: & = \: \frac{\rho f_x}{2 \mu} \: \left( \frac{\rho}{\kappa} c_p \frac{\partial T}{\partial x} + 2 \frac{\rho f_x}{\kappa} \right) \frac{h^3}{24} \left( L_K +\frac{h}{8} \right) \\
& - \frac{\rho f_x}{2 \mu} \frac{\rho}{\kappa} c_p \frac{\partial T}{\partial x} \left( \frac{h^2}{4} + hL_s \right) \frac{h}{2} \left( L_K +\frac{h}{4} \right)
\end{align*}
where $L_K$ is the Kapitza length.
In \Fig{temperature_developing} it is shown that temperature profiles may indeed be very well fitted by this solution. At the beginning of the channel the convective term dominates and the profile is eminently quadratic; on the contrary, near the end where the flow is almost thermally developed $a_2 \approx 0$ and the profile is $\propto y^4$.

\begin{figure}[h!]		
	\begin{center}
		\includegraphics[width=0.7\textwidth]{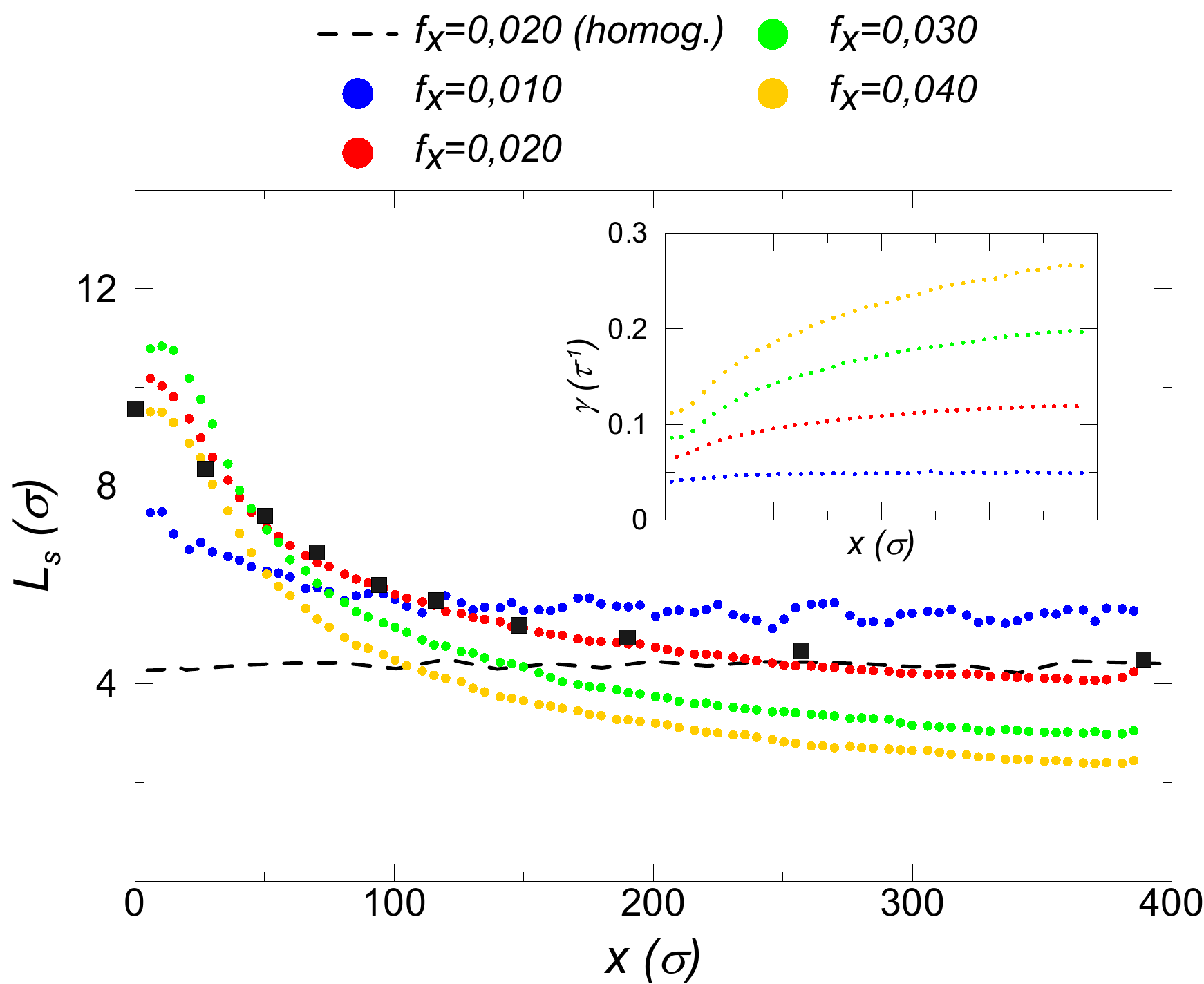}
		\caption{Slip length calculated along the flow direction for the SH (black dashed curve) and SIFD-CR models with different $f_x$. In the inset, the shear rate distributions along the channel for different values of $f_x$ are shown. The squared black symbols correspond to simulations in which the fluid is thermostated (see text).}
		\label{fig-slip_length_vs_x}
	\end{center}	
\end{figure}

Finally, we have focused on the results obtained for the flow slip over the solid surface. The study of the slip observed at microscales and nanoscales remains to be of great interest at present, among other reasons, because of its potential technological utility for nanoscale flows: since the occurrence of slip in nanochannels reduces the fluid-solid friction and increases the flow rate, it is a phenomenon sought in many nanofluidic applications \citep{Bocquet2010,Holt2006}. Therefore, being able to establish a specific boundary condition for fluid flows over solid surfaces would be fundamental for understanding flows in these scales. But, although this phenomenon has been extensively investigated from experimental, theoretical and computational points of view \citep{Lauga2007,Neto2005}, there are some issues that are still controversial, like the slip dependence on shear rate. Since the seminal work of \cite{Thompson1997}, some authors have reported (both in experimental and simulation studies) a non-bounded monotonic increase of the slip length with shear rate and the existence of a critical shear rate at which $L_s$ diverges \citep{Thompson1997,Zhu2001,Priezjev2007a}. However, other researchers have found that slip length 
tends to a finite constant value at high shear \citep{Martini,AlizadehPahlavan2011}. It is important to emphasize that in all these works an homogeneous flow is assumed, the first group applying a thermostat to the fluid and the second only to the solid \citep{AlizadehPahlavan2011}.

In \Fig{slip_length_vs_x} we can see the slip length for the SIFD flow model, calculated from definition in \Eq{slip_length} and the velocity profiles obtained in MD simulation. The slip length is higher at the inlet, decreases along $x$ and tends to a constant value when the flow is hydrodynamically developed (between $2 \: \sigma$ and $6 \: \sigma$ in the simulated range of forces). It is worth pointing out that, for $f_x = 0.020 \: \epsilon / m\sigma$, the slip length value at high $x$ approximately coincides with that obtained with the homogeneous model and the same force. We confirm again, then, that the study of homogeneous flow can describe the developed flow, but not its developing behavior. We also see that shear rate increases along the channel, as it can be readily understood from the increasing slope of velocity profiles at the boundaries in \Fig{velocity_developing} (see inset). The fact that slip reduces with growing shear rate could appear to be in contradiction with those works that, in the line of \cite{Thompson1997}, conclude that slip grows with shear. However, those works assume constant temperature. Temperature variation affects the slip, as has already been highlighted in the literature \citep{Guo2005}: when $T$ increases, fluid particles become more active and a higher number of them are able to penetrate in the region of interaction with the wall, in such a way that the momentum transfer between the fluid and the solid improves, and the velocity slip between them decreases \citep{Guo2005}. We suggest, therefore, that the evolution of slip in the channel is due to the rise in temperature in it. In order to support this conclusion, we have carried out ten extra MD simulations of the homogeneous model with $f_x = 0.020 \: \epsilon / m\sigma$, but now thermostating also the fluid (with a DPD thermostat) to force the fluid temperature and density to be equal to those at ten different sections of the channel. The results, shown with squared symbols in \Fig{slip_length_vs_x}, confirm that temperature is the crucial factor that makes the slip to decrease along the channel, even if the shear rate gradually increases. It also explains the smaller $L_s$ for higher $f_x$ (notice that at lower forces, thermal fluctuations cause noisier results, and additional averaging would be needed to smooth them).

\subsection{Open or closed reservoirs?}

\begin{figure}[h!]		
	\begin{center}
		\includegraphics[width=0.7\textwidth]{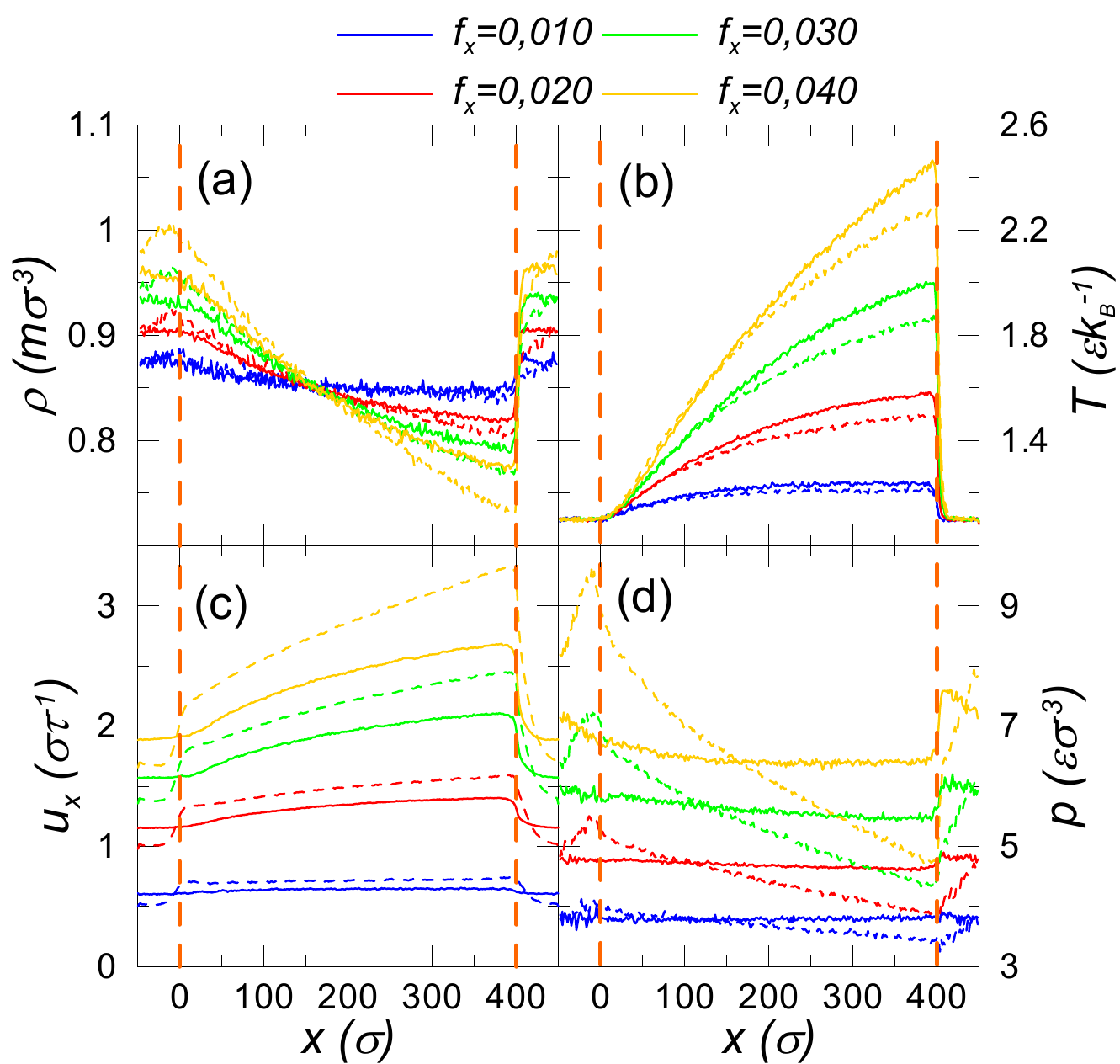}
		\caption{Averaged fluid profiles in the plane $y=0$ along the flow direction, obtained with the SIFD model: (a) density, (b) temperature, (c) streaming velocity, and (d) pressure. Solid and dashed curves correspond to closed reservoirs (SIFD-CR) and open reservoirs (SIFD-OR) model, respectively.}
		\label{fig-open_vs_closed_x}
	\end{center}	
\end{figure}

\begin{figure}[h!]		
	\begin{center}
		\includegraphics[width=0.4\textwidth]{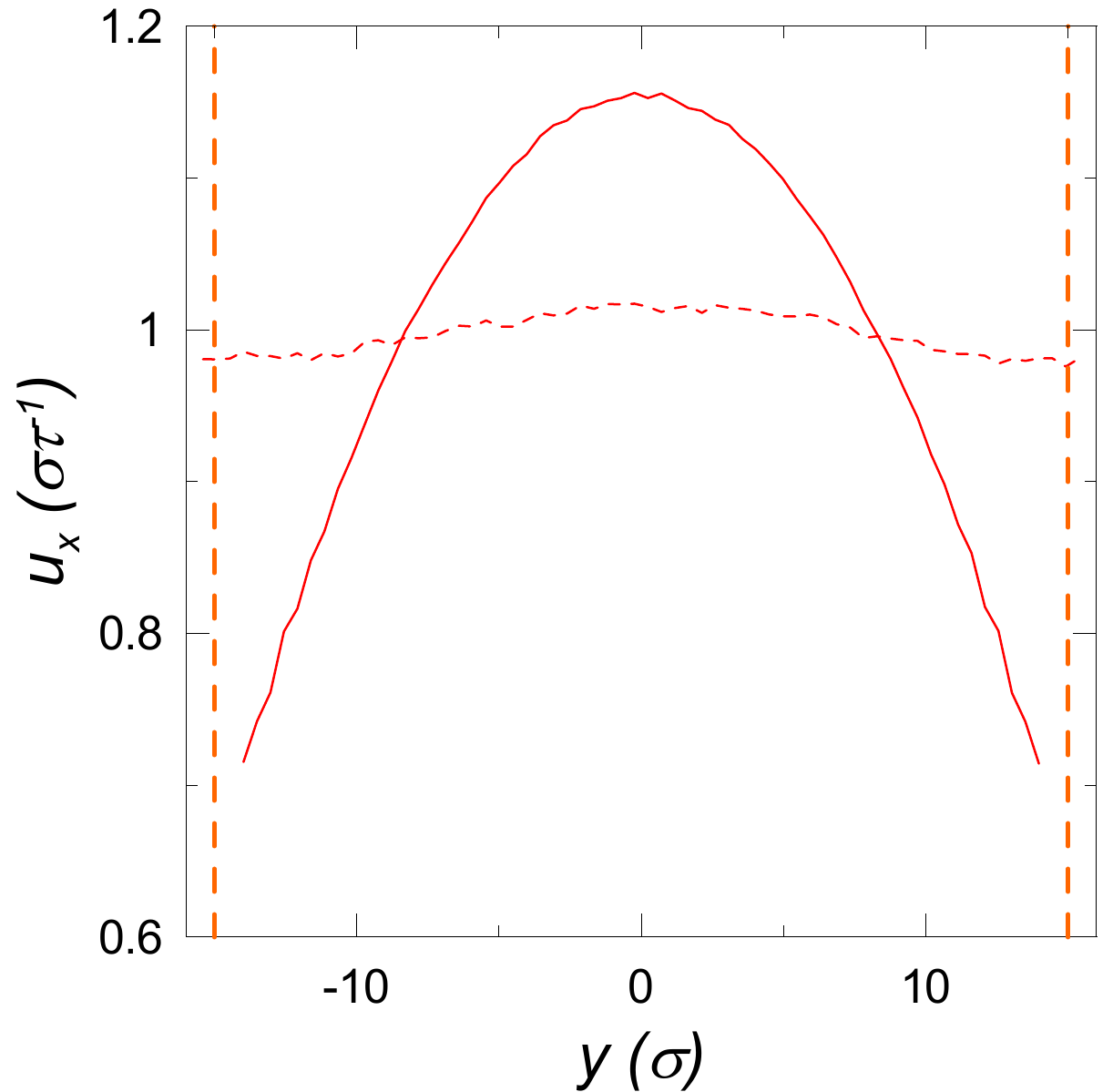}
		\caption{Velocity distribution across the channel in reservoir region ($x = -25 \: \sigma$) with $f_x = 0.02 \: \epsilon / m\sigma$. Solid and dashed curves correspond to closed reservoirs (SIFD-CR) and open reservoirs (SIFD-OR) model, respectively.}
		\label{fig-reservoir_velocity}
	\end{center}	
\end{figure}

A few comments on the effects of assuming open reservoirs outside the channel (configuration in \Fig{channel-developing} instead of that in \Fig{channel-developing2}) should also be made, since they can shed some light on the discussion about boundary conditions choice in MD simulations of nanoflows. The main difference with respect to the closed-reservoirs case reported so far lies on the pressure gradient created out of the channel (see \Fig{open_vs_closed_x}(d)). The lack of walls in the open reservoirs causes much flatter velocity profiles (in the reservoir) than those in the closed ones, as shown in \Fig{reservoir_velocity}, and then, much smaller viscous forces in these regions. As a result, a positive pressure gradient appears to compensate the external force (see \Eq{momentum}). This pressure difference between channel ends translates in a pressure drop inside the channel, and in an extra force on the confined fluid which adds to $f_x$. Its effects are not minor, since $-\partial p / \partial x$ is comparable to $\rho f_x$. It must therefore be concluded that simulated systems with the same force but different boundary conditions may not be dynamically equivalent, and this must be taken into account when designing the model to simulate.

As a consequence a higher flow rate is observed when open reservoirs are considered, as it can be seen from \Fig{open_vs_closed_x}(a) and \Fig{open_vs_closed_x}(c) (note that, for example at $x\simeq 150 \sigma$, the densities are similar but the velocity is bigger when open reservoirs are considered). Evidently, the hydraulic resistance of closed reservoirs is higher.

On the other side, although the higher force exerted on the confined fluid in the open-reservoirs configuration (and the corresponding higher shear rate) could suggest a more intensive heating, the temperature distribution along the channel is not substantially different (and shows even a lower $T$) from the closed-reservoirs case (see \Fig{open_vs_closed_x}(b)). The explanation for this behavior can be found in the second term of the left-hand side of the energy equation, \Eq{energy}: a fraction of the heat transferred to the fluid is devoted to increase the fluid temperature, but another part goes to diminish the fluid pressure (unlike what happens with closed reservoirs). Also the variation of shear viscosity in the flow direction could affect $T$ distribution at some extent (see a detailed discussion of this issue at the next subsection). 

To conclude this subsection we also note that {\color{red} in the case of closed reservoirs,} caution is recommended when choosing the reservoirs height (in $y$ direction), since it can affect the results in some measure. The reason is that, for a given channel width $L_y$, increasing the reservoirs height (and then the wall thickness $\Delta y_w$) results in bigger pressure losses at the entrance (since the flow contraction is more abrupt), which, in turn, results in
%and the friction at vertical solid walls more significant
%and then yields 
a reduction in the pressure gradient inside the channel. This influences the fluid properties obtained because, as discussed by \cite{Thomas2009}, it is this pressure gradient, and not the pressure difference between reservoirs, which characterizes the flow (see also \cite{Nichollsa}). We have checked that entrance losses increase indeed if reservoirs height is enlarged, but it affects only slightly the presented results. 

\subsection{Streamwise inhomogeneous pressure-driven flow}

Finally, we now move to discuss the third and last type of models studied in this work, which should be, a priori, the most realistic to simulate nanoflows, since its driving mechanism is not a fictitious external force that disturbs significantly the behavior of the fluid inside the channel, but a pressure gradient (obviously, induced also by a force but applied in this case far enough from the channel).

\begin{figure}[h!]
	\begin{center}
		\includegraphics[width=0.7\textwidth]{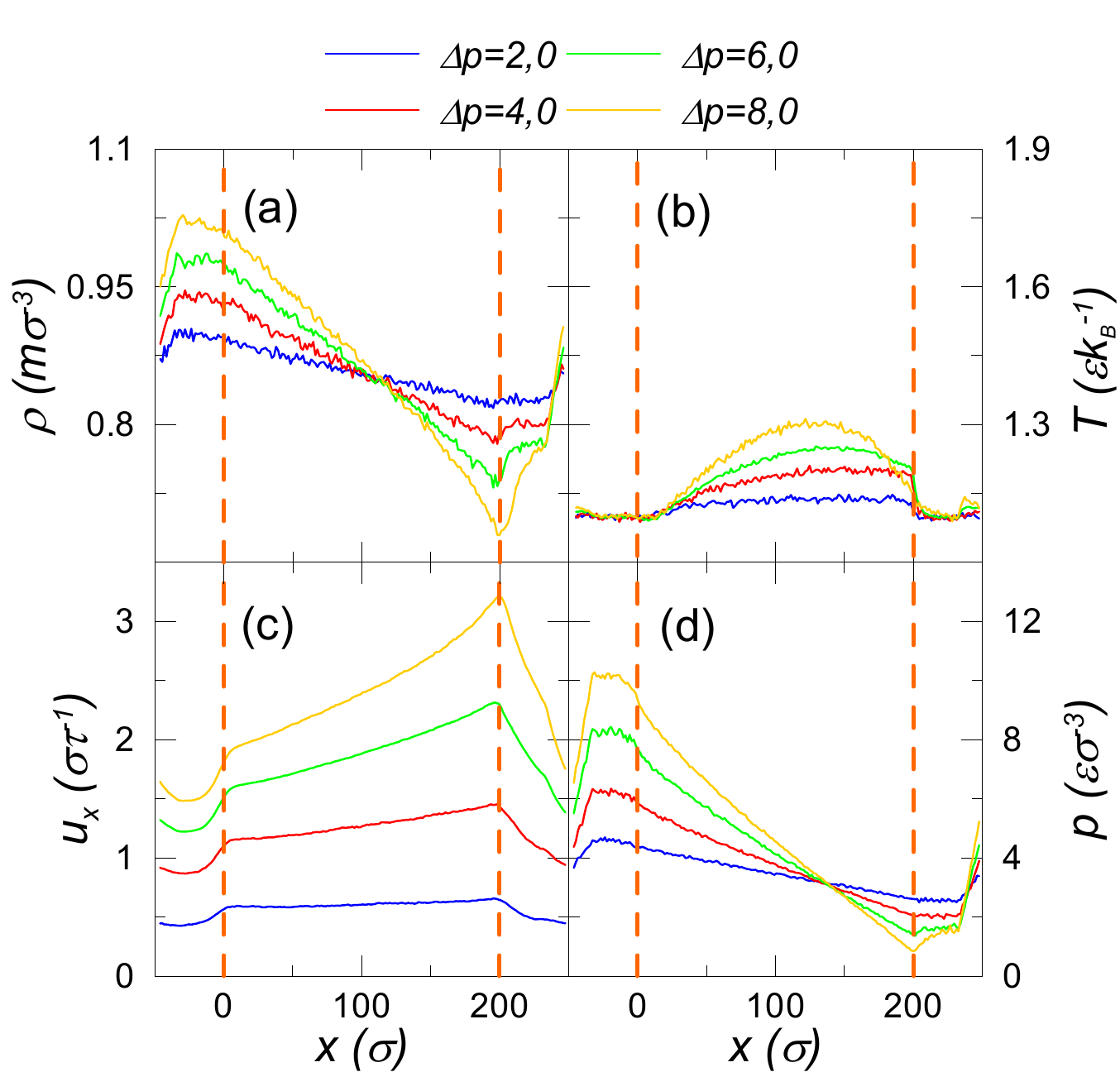}
		\caption{Averaged fluid profiles in the plane $y=0$ along the flow direction, obtained with the SIPD model: (a) density, (b) temperature, (c) streaming velocity, and (d) pressure.}
		\label{fig-pressure_driven_streamwise_profiles}
	\end{center}	
\end{figure}

Firstly, it has been confirmed that the application of a force of the magnitude in \Eq{force_model3} in the margin regions of length $L_{res}/3$ (see \Fig{channel-developing}) translates in a pressure difference between the ends of the channel which coincides with the $\Delta p$ value imposed in \Eq{force_model3} with satisfactory accuracy (less than a 10\% discrepancy). In \Fig{pressure_driven_streamwise_profiles}(d) we present the pressure profiles for a channel of length $L_x = 200 \: \sigma$ and four different $\Delta p$ values, chosen to create the same driving in the confined fluid as the one in the SIFD-OR model shown in \Fig{open_vs_closed_x} (that is, the value of $\Delta p$ in the SIPD model is chosen such that $\Delta p / L_x$ in this model equals the average driving $ \Delta p / L_x + \rho_ff_x$ in the SIFD-OR model for each value of $f_x$ shown in \Fig{open_vs_closed_x}). This choice aims to compare dynamically equivalent flows. %In so doing we are allowed to compare gravity- to pressure-driven flows, since they are dynamically equivalent. 
Observe that in the SIPD model the induced pressure gradient is much bigger than in the SIFD one. As it will be shown throughout this section, the pressure variation affects the rest of thermodynamic fluid properties and changes notably the results analyzed so far. Also note that pressure losses at the channel entrance, between the point where the external force is no longer applied and the inlet at $x=0$, are barely appreciable.

\begin{figure}[h!]		
	\begin{center}
		\includegraphics[width=0.7\textwidth]{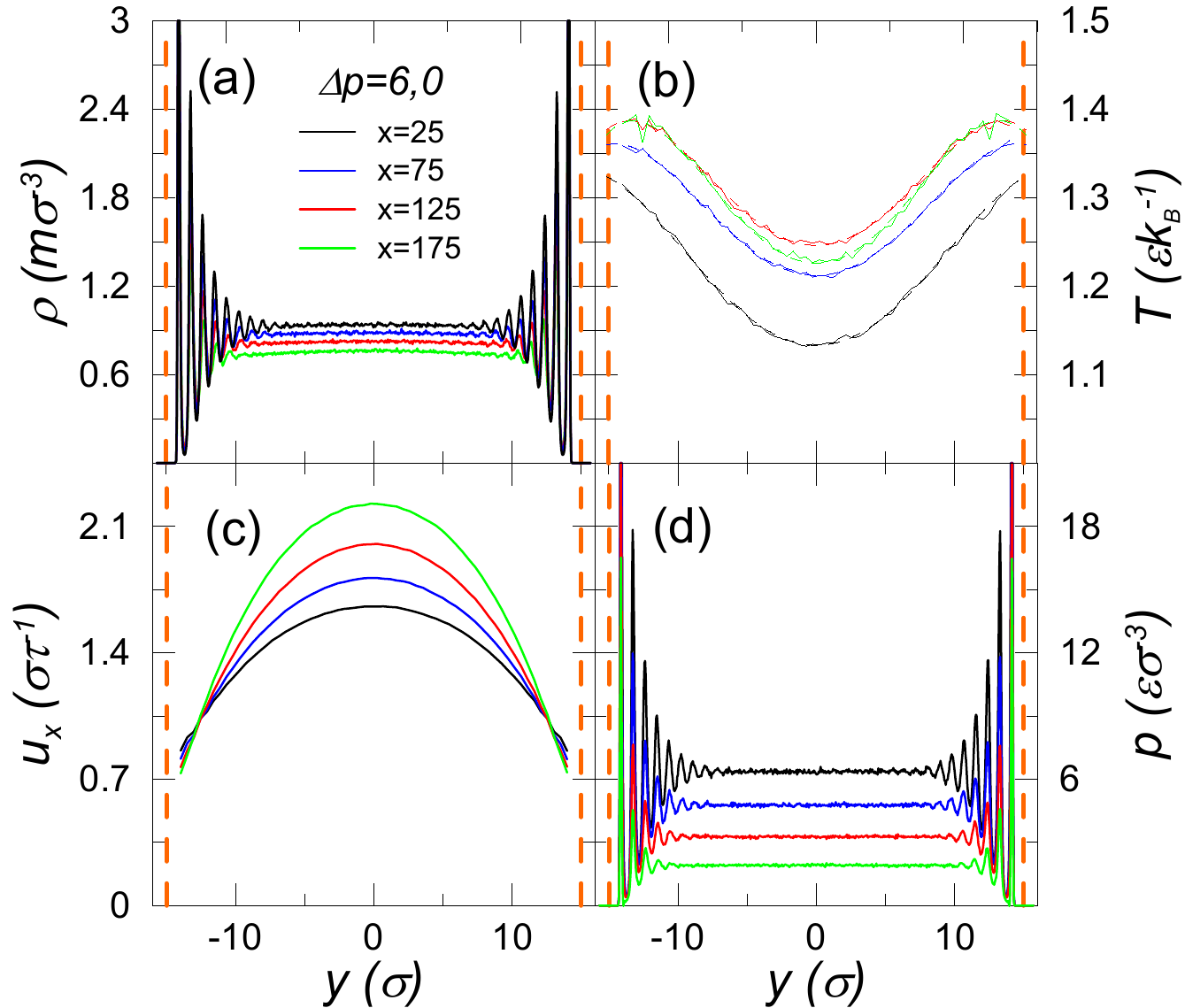}
		\caption{Averaged fluid profiles across different sections of a channel of length $L_x = 200 \: \sigma$, obtained with the SIPD model and a pressure difference $\Delta p = 6.0 \: \epsilon / \sigma^3$: $x=25 \: \sigma$ (black curves), $x=75 \: \sigma$ (blue curves), $x=125 \: \sigma$ (red curves), and $x=175 \: \sigma$ (green curves): (a) density, (b) temperature, (c) streaming velocity, and (d) pressure. Dashed curves in (b) are the best fits for temperature profiles with the form in \Eq{temperature_quartic+quadratic}.}
		\label{fig-pressure_driven_wall_normal_profiles}
	\end{center}	
\end{figure}

\begin{figure}[h!]		
	\begin{center}
		\includegraphics[width=1.0\textwidth]{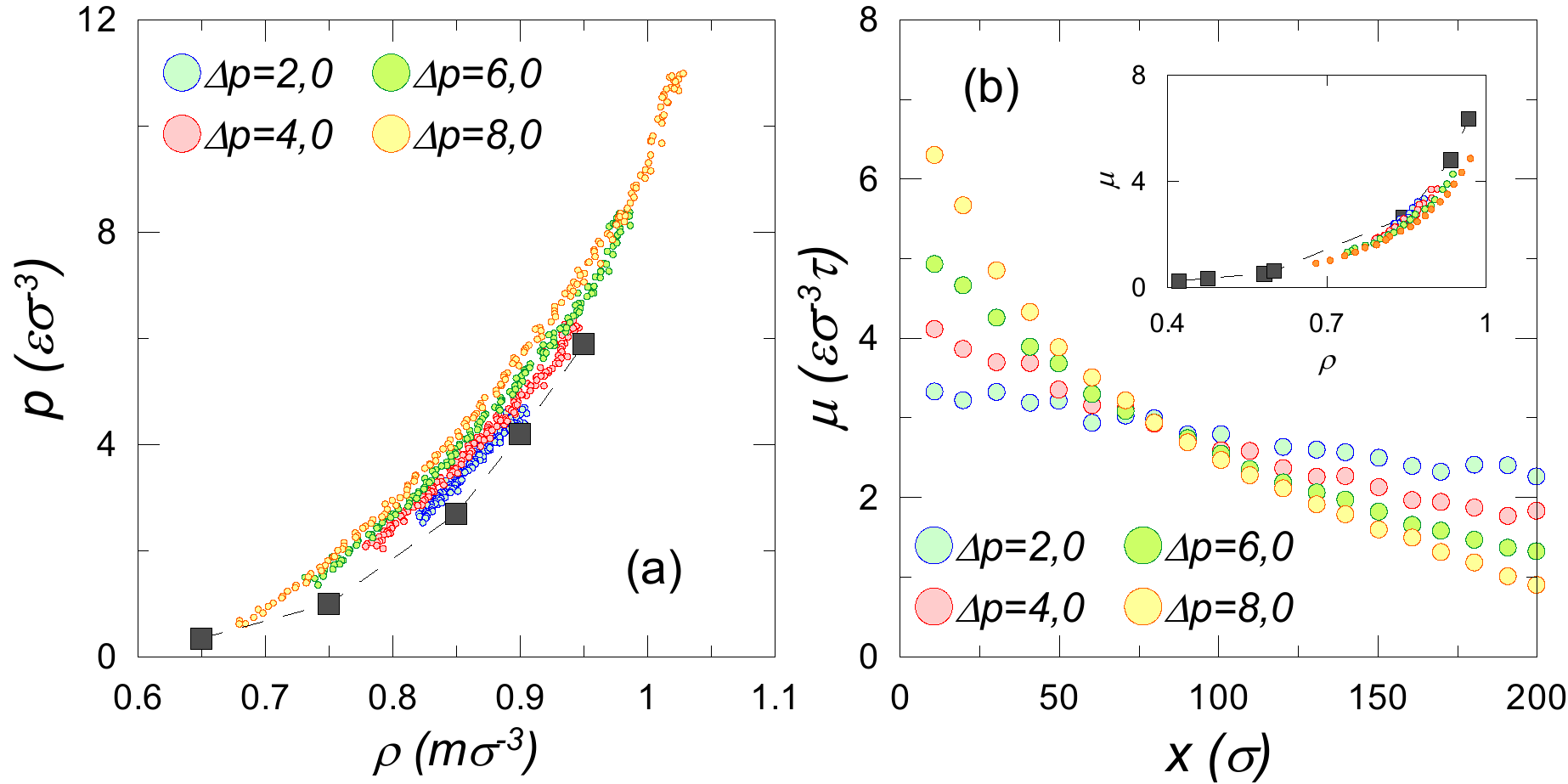}
		\caption{Left panel: Pressure versus fluid density obtained for various pressure differences $\Delta p$. Black squares are extracted from the phase diagram reported by \cite{Nicolas1979} from MD simulations. Right panel: Viscosity along the channel for different $\Delta p$. In the inset we compare the evolution of $\mu$ with fluid density in these same simulations to the MD results of \cite{Tankeshwar2000a} for a temperature of 1.23 $\epsilon / K_B$.}
		\label{fig-pressure_viscosity_exp}
	\end{center}	
\end{figure}

The induced pressure difference causes a significantly more pronounced variation of the density along the channel than in the SIFD models with the same total force (see \Fig{pressure_driven_streamwise_profiles}(a)). In fact this density variation limits the applicability of this kind of models, since if the pressure drop is too high, $\rho$ will diminish sufficiently to provoke a phase change at the exit of the channel. This imposes a limitation on the maximum $\Delta p$ applied in MD simulations, and on the channel length for a given pressure gradient. This is the reason why we are reporting results only for $L_x = 200 \: \sigma \:$: simulations with larger $L_x$ demand also larger $\Delta p$ to induce a certain gradient, and the phase transition occurs. Also related to density variation, it should be noted that the profiles in the wall-normal direction ($y$) show a more marked structure near the walls (with more clearly located atomic layers) at the beginning of the channel, where density is higher, as is apparent in \Fig{pressure_driven_wall_normal_profiles}(a) and \Fig{pressure_driven_wall_normal_profiles}(d). On the other hand, it has been verified that in our simulations fluid density evolves with pressure in a qualitatively similar fashion to that reported by \cite{Nicolas1979}, who obtained the phase diagram of a Lennard-Jones fluid at equilibrium by MD simulations. As it can be seen in \Fig{pressure_viscosity_exp}(a), for small $\Delta p$ the $p$-$\rho$ relation approaches the equilibrium equation of state, while for larger $\Delta p$ the pressure is slightly higher than the one at equilibrium but the functional relation with the density is similar.

\begin{figure}[h!]		
	\begin{center}
		\includegraphics[width=0.5\textwidth]{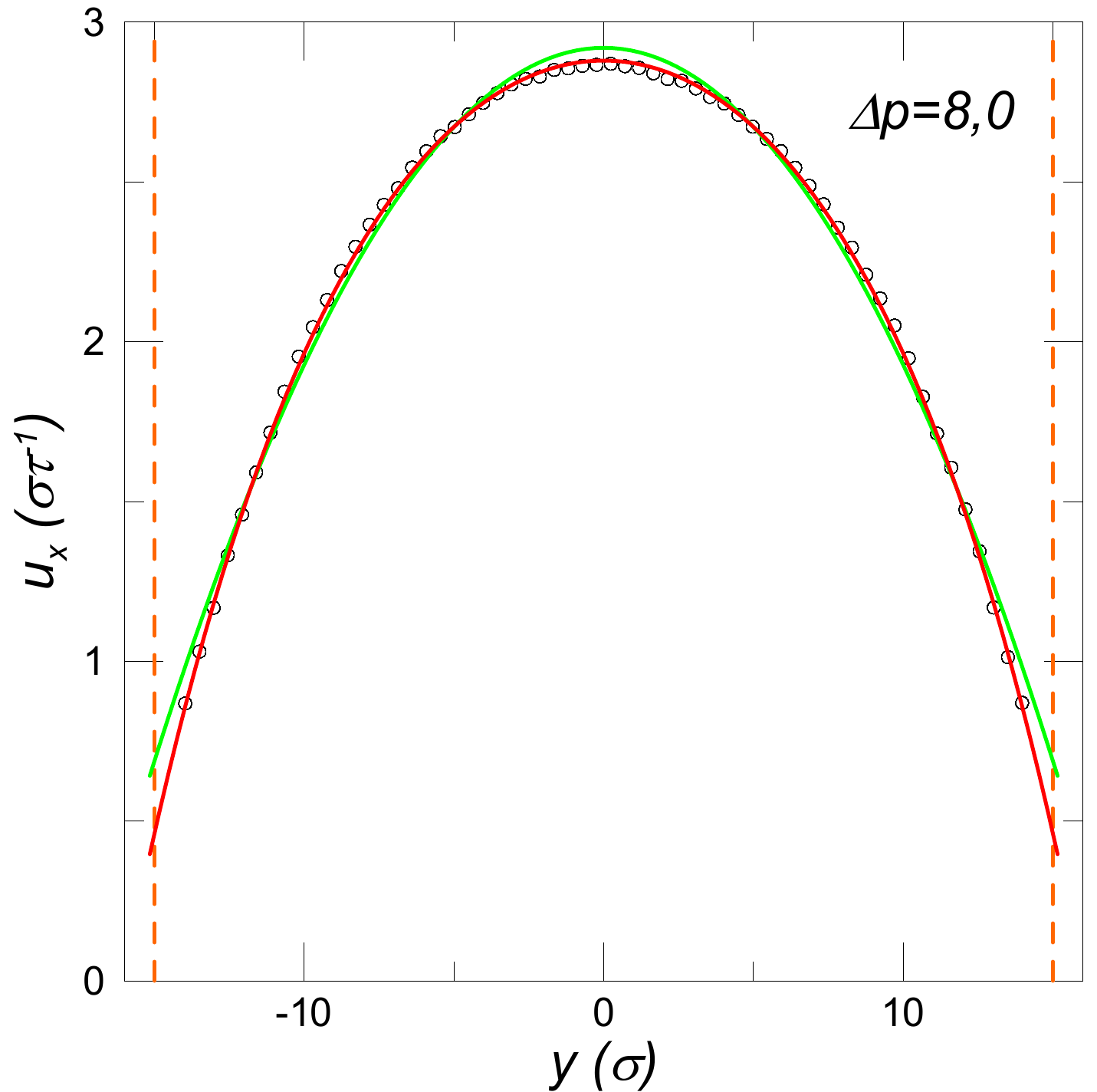}
		\caption{Open symbols: Velocity profile across section $x = 175 \: \sigma$ obtained with a pressure-driven model with $\Delta p = 8,0 \: \epsilon / \sigma^3$ and $L_x = 200 \: \sigma$. Solid curves are the best quadratic  (green curve, eq. \Eq{velocity_quadratic}) and hyperbolic (red curve, \Eq{velocity_hypcosin}) velocity fits.}
		\label{fig-pressure_driven_vel_fitting}
	\end{center}	
\end{figure}

The averaged velocity also shows a faster growth in the channel when flow is induced by a difference in pressure (as it can be seen comparing \Fig{pressure_driven_streamwise_profiles}(c) with \Fig{open_vs_closed_x}(c), which is consistent with the greater density drop and the requirement of mass flow rate conservation along the channel. We can also observe that the gradient of $u_x$ progressively increases along $x$, and it is clearly larger at the exit; that is, the fluid is more accelerated near the end than at the beginning of the channel. Besides, this effect is more pronounced for higher $\Delta p$, in fact it is hardly noticeable for $\Delta p = 2.0$ but clearly visible for $\Delta p = 8.0$. One might ask for the physical cause of this behavior. Since pressure gradient does not change appreciably along $x$, we suggest that, again, it is the intense change of fluid properties in the channel (in this case, shear viscosity) which explains it. Figure~\ref{fig-pressure_viscosity_exp}(b) includes the results for viscosity as a function of $x$ for different $\Delta p$, extracted from the simulated shear stress through $P_{xy} \: = \: \mu \left( \partial u_x / \partial y \right)$. $\mu$ lowering along $x$ is in fact significant, being more marked as pressure gradient is increased. This tendency is consistent with the results of \cite{Tankeshwar2000a}, who reported both theoretical and MD calculations for shear viscosity at a wide range of temperatures, and showed that $\mu$ diminishes when $\rho$ decreases (see the inset in \Fig{pressure_viscosity_exp}(b)). This behavior indicates that friction is reduced along the channel, and then explains the increase in the gradient of $u_x$. Precisely at those regions where $\partial u_x / \partial x$ grows, the hyperbolic function \Eq{velocity_hypcosin} starts to differ from the quadratic function \Eq{velocity_quadratic}, and one can confirm that it is more suitable to fit the velocity profiles, although the discrepancy is still small (as an example, see the velocity in a point near the end of the channel for $\Delta p = 8.0 \: \epsilon / \sigma^3$ in \Fig{pressure_driven_vel_fitting}).

But certainly the most significant difference observed in our MD simulations between the SIFD and the SIPD flow models resides in the temperature distribution along the channel. If we look at \Fig{pressure_driven_streamwise_profiles}(b), we clearly observe that in SIPD models $T$ raises to a much lesser extent than in SIFD models (\Fig{open_vs_closed_x}(b)). The difference is important enough to conclude that the choice of proper boundary conditions is a fundamental question in MD simulations of nanoflows, and must be addressed carefully. In this work we suggest that the cause of this disparity in the evolution of $T$ is twofold. In the first place, as we mentioned for the case of models with an external force and open reservoirs, the term $- \beta T u_x \frac{\partial p}{\partial x}$ in the energy equation, \Eq{energy}, acts as an effective cooling mechanism. The internal energy increase produced by the viscous heat does not directly result in a temperature increase, as it would occur in an incompressible flow, due to the energy required by the pressure loss to occur. As the flow velocity $u_x$ increases along the channel, this contribution becomes higher and temperature growth becomes progressively slower (for $\Delta p$ larger than $6.0 \epsilon / \sigma^3$ one can even observe a slight $T$ reduction at the end of the channel). The second factor that contributes to moderate the temperature is shear viscosity, that, as we have seen, decreases in the flow direction, then causing a gradual reduction of the friction. The relative importance of these two causes is not clear, and deserves further research. What we do know is that both become much more important in models in which flow is induced by a pressure gradient, since the streamwise pressure and viscosity variation increases notably with respect to those driven by a uniform external force. With regard to the form of temperature distribution across the flow (\Fig{pressure_driven_wall_normal_profiles}(b)), we see again that $T$ profiles meet the functional form derived in \Eq{temperature_quartic+quadratic}. Compared to those presented in \Fig{temperature_developing} for SIFD flow models, the quadratic term (we recall that it vanishes for SH flow models) dominates over the fourth-order term. 

\section{Conclusions}
\label{sec-conclusions}
A careful analysis of three different MD models for the flow in nanochannels has been reported. The traditional SH (force driven) flow model, which does not account for the variation of properties along the channel, permits to predict the density, temperature, pressure and velocity profiles (and thus slip length) when low forces are applied. In this case the heat generation by friction is small and it is easily dissipated by thermal conduction to the walls, where it is finally dissipated by the thermostat applied there. Other heat transfer (cooling) mechanisms, convection and dilatation, are missing as they are incompatible with an homogeneous flow. Therefore, at higher forces an inhomogeneous model must be used to capture flow developing profiles if the associated computational cost can be afforded.

When two reservoirs are added at the inlet and the outlet and the fluid is thermostated there to fix the inlet temperature, streamwise variation of the flow can be predicted. The main difficulty here is that the results depend on the design of the reservoirs. If the reservoirs are surrounded by (fictitious) extensions of the walls no pressure gradient is generated because the flow is driven by an external force that balances the viscous dissipation, in the same way as it occurs inside the channel. If open reservoirs are considered, the velocities outside the channel are almost uniform and a pressure gradient is generated. In the former case, we have seen that also the reservoirs size can affect the pressure distribution inside the channel, although the influence in the results presented in this work is minor.

The inclusion of reservoirs outside the domain of interest allows us to analyze the streamwise evolution of the shape of velocity and temperature profiles in the wall-normal direction. In particular, the appearance of a quadratic term in $T(y)$ as a consequence of convection is discussed. For the higher forces in the range studied in this work, the flow is not fully (hydrodynamically and thermally) developed at the end of the channel, despite the large simulated channel lengths. The usual homogeneous simulations ignore this developing behavior, as well as the stabilization of the slip length along the channel.

The pressure gradient inside the channel has an important influence on the results. Even if the pressure profile were constant across the channel section, a pressure gradient is equivalent to a constant external force only for incompressible flows. At high pressure differences, heat generation makes compressibility effects important, the density cannot be assumed to be constant and the dilatation work acts as a heat sink. Therefore, the results obtained using the SIPD model are substantially different from those obtained using the SIFD model, specially regarding temperature distribution. It has been demonstrated that the temperature growth along the channel is much smaller in SIPD than in SIFD models. Both the energy required by the pressure loss and the streamwise variation of viscosity are identified as the factors which explain this behavior.

In this respect it is worth noting that if a low cost SH model is to be used, thermostating the fluid to account for the missing heat transfer mechanisms will produce better (but still inaccurate) results than those obtained with the SH model in which only walls are thermostated. This conclusion is obtained comparing the temperature distributions in \Fig{open_vs_closed_x} and \Fig{pressure_driven_streamwise_profiles}: the temperature obtained with the most realistic model (SIPD) do not exceed $1.3\: \epsilon/k_B$ whereas the temperatures obtained using the SH model almost doubles this value. It is therefore less inaccurate to consider the temperature fixed at its inlet value ($1.1\: \epsilon/k_B$). Nevertheless, we remark once again that the use of homogeneous models will only provide a first approximation due to their inability to describe the streamwise variation of flow properties.

\bibliographystyle{abbrv}
\bibliography{art-2d-preprint}

\end{document}